\documentclass[aps,prd,twocolumn,superscriptaddress]{revtex4}
\usepackage{epsfig,epsf}
\usepackage{amsmath}
\usepackage{amsthm}
\usepackage{amsfonts}
\usepackage{amssymb}
\usepackage{dsfont}
\usepackage{multirow}
\usepackage{appendix}
\usepackage{slashed}
\usepackage[active]{srcltx}
\usepackage{psfrag}

\begin{document}

\title{Newly discovered $\Xi_c^{0}$ resonances and their parameters}
\date{\today}
\author{S.~S.~Agaev}
\affiliation{Institute for Physical Problems, Baku State University, Az--1148 Baku,
Azerbaijan}
\author{K.~Azizi}
\affiliation{Department of Physics, University of Tehran, North Karegar Avenue, Tehran
14395-547, Iran}
\affiliation{Department of Physics, Do\v{g}u\c{s} University, Acibadem-Kadik\"{o}y, 34722
Istanbul, Turkey}
\affiliation{School of Particles and Accelerators, Institute for Research in Fundamental
Sciences (IPM) P.O. Box 19395-5531, Tehran, Iran}
\author{H.~Sundu}
\affiliation{Department of Physics, Kocaeli University, 41380 Izmit, Turkey}

\begin{abstract}
The aim of the present article is investigation of the newly observed
resonances $\Xi_c(2923)^{0}$, $\Xi_c(2939)^{0}$, and $\Xi_c(2965)^{0}$ which
are real candidates to charm-strange baryons. To this end, we calculate the
mass and pole residue of the ground-state and excited $1P$ and $2S$ spin-$%
1/2 $ flavor-sextet baryons $\Xi_{c}^{\prime 0}$, $\Xi_{c}^{\prime
0}(1/2^{-})$ and $\Xi_{c}^{\prime 0}(1/2^{+})$ with quark content $csd$,
respectively. The masses and pole residues of the ground-state and excited
spin-$3/2$ baryons $\Xi_{c}^{\star 0}$ are found as well. Spectroscopic
parameters of these particles are computed in the context of the QCD
two-point sum rule method. Widths of the excited baryons are evaluated
through their decays to final states $\Lambda_{c}^{+}K^{-}$ and $%
\Xi_{c}^{\prime 0}\pi$. These processes are explored by means of the full
QCD light-cone sum rule method necessary to determine strong couplings at
relevant vertices. Obtained predictions for the masses and widths of the
four excited baryons, as well as previous results for $1P$ and $2S$
flavor-antitriplet spin-1/2 particles $\Xi_c^{0}$ are confronted with
available experimental data on $\Xi_{c}^{0}$ resonances to fix their quantum
numbers. Our comparison demonstrates that the resonances $\Xi_c(2923)^{0}$
and $\Xi_c(2939)^{0}$ can be considered as $1P$ excitations of the spin-$1/2$
flavor-sextet and spin-$3/2$ baryons, respectively. The resonance $\Xi
_{c}(2965)^{0}$ may be interpreted as the excited $2S$ state of either spin-$%
1/2$ flavor-sextet or antitriplet baryon.
\end{abstract}

\maketitle

%

\section{Introduction}

\label{sec:Introduction}
The discovery of three new resonances $\Xi _{c}(2923)^{0}$, $\Xi
_{c}(2939)^{0}$, and $\Xi _{c}(2965)^{0}$ by the LHCb collaboration is a
last result of the experiments devoted to investigation of charmed and
bottom baryons with different spin-parities and quark contents \cite%
{Aaij:2020yyt}. Five narrow states $\Omega _{c}^{0}$ fixed in the $\Xi
_{c}^{+}K^{-}$ invariant mass distribution \cite{Aaij:2017nav}, and four
peaks $\Omega _{b}^{-}$ detected recently in the $\Xi _{b}^{0}K^{-}$
spectrum \cite{Aaij:2020cex} were results of previous measurements performed
by LHCb.

Needless to say, that discovery of the resonances $\Omega _{c}^{0}$
stimulated numerous studies of excited charmed baryons aimed to understand
their internal organizations and quantum numbers. Actually, heavy flavored
baryons were already objects of theoretical analyses, in which spectroscopic
parameters of the ground state and excited particles, their decay channels
and strong couplings, magnetic moments and radiative decays were studied by
means of different models and methods of high energy physics. New
experimental information on $\Omega _{c}^{0}$, besides traditional models,
gave rise to their interpretations as exotic pentaquark states. In our
articles \cite{Agaev:2017jyt,Agaev:2017lip,Agaev:2017ywp}, we investigated
the baryons $\Omega _{c}^{0}$ and $\Omega _{b}^{-}$, where one can find
further details and references to relevant publications.

The baryons from the $\Xi _{c}^{0}=csd$ family are another interesting
objects for both experimental and theoretical analyses. Parameters of the
ground-state $J^{\mathrm{P}}=1/2^{+}$ and $3/2^{+}$ baryons with the content
$csd$ were measured already and included into Particle Data Group (PDG)
tables \cite{Tanabashi:2018oca}. Thus, the mass and mean lifetime of the
flavor-antitriplet baryon $\Xi _{c}^{0}$ are
\begin{equation}
m_{\mathrm{exp}}=2470.90_{-0.29}^{+0.22}~\mathrm{MeV},\,\,\tau _{\mathrm{exp}%
}=(1.53\pm 0.06)\times 10^{-13}~\mathrm{s},  \label{eq:Data1}
\end{equation}%
whereas for the mass of the flavor-sextet $J^{\mathrm{P}}=1/2^{+}$
ground-state particle $\Xi _{c}^{\prime 0}$ we have%
\begin{equation}
m_{\mathrm{exp}}=(2579.2\pm 0.5)~\mathrm{MeV}.  \label{eq:Data2}
\end{equation}%
The mass of the $J^{\mathrm{P}}=3/2^{+}$ baryon $\Xi _{c}(2645)^{0}$ is also
known
\begin{equation}
m_{\mathrm{exp}}^{\ast }=2645.56_{-0.30}^{+0.24}~\mathrm{MeV}.
\label{eq:Data3}
\end{equation}%
There are a few charged and neutral particles of this family listed in Ref.\
\cite{Tanabashi:2018oca}, some of which will be considered in the last
section of the present work.

As we have noted above, theoretical investigations of heavy flavored
baryons, including $\Xi _{c}$ ones, have long history \cite%
{Capstick:1986bm,Ebert:2007nw,Ebert:2011kk,Garcilazo:2007eh,Valcarce:2008dr,Roberts:2007ni,Yoshida:2015tia, Shah:2016nxi,Bagan:1992tp,Huang:2000tn,Wang:2002ts,Wang:2009cr,Chen:2015kpa,Chen:2016phw,Chen:2017sci, Aliev:2010yx,Aliev:2010ev,Aliev:2011ufa,Azizi:2015ksa,Chiladze:1997ev,Edwards:2012fx,Padmanath:2013bla}%
. These particles were explored in the context of various quark models \cite%
{Capstick:1986bm,
Ebert:2007nw,Ebert:2011kk,Garcilazo:2007eh,Valcarce:2008dr,Roberts:2007ni,Yoshida:2015tia,Shah:2016nxi}%
, by using the QCD sum rule method \cite%
{Bagan:1992tp,Huang:2000tn,Wang:2002ts, Wang:2009cr,
Chen:2015kpa,Chen:2016phw,Chen:2017sci,Aliev:2010yx,Aliev:2010ev,Aliev:2011ufa,Azizi:2015ksa}%
, by means of the Heavy Quark Effective Theory (HQET) \cite{Chiladze:1997ev}
and lattice simulations \cite%
{Edwards:2012fx,Padmanath:2013bla,Bahtiyar:2020uuj}.

The discovery of three new resonances by LHCb added valuable knowledge about
excited baryons $\Xi _{c}^{0}$, which together with $\Xi _{c}(2930)^{0}$ and
$\Xi _{c}(2970)^{0}$ generated theoretical activities to explain their
parameters. Problem is that LHCb did not inform on spins and parities of
these resonances, which are important topic of continuing theoretical
studies. Here, it is necessary to give some information about the resonance $%
\Xi _{c}(2930)^{0}$, which is relatively "old" member of this family. It was
observed by the BaBar collaboration as the intermediate resonant structure
in the process $B^{-}\rightarrow \Lambda _{c}^{+}\bar{\Lambda}_{c}^{-}K^{-}$
\cite{Aubert:2007eb}. Existence of $\Xi _{c}(2930)^{0}$ was confirmed
recently by Belle in Ref.\ \cite{Li:2017uvv}, in which the collaboration
reported about its observation as a resonance in the $\Lambda _{c}^{+}K^{-}$
invariant mass spectrum in the same decay process. The mass and width of
this state reported by Belle are
\begin{eqnarray}
m &=&(2928.9\pm 3.0_{-12.0}^{+0.9})~\mathrm{MeV},  \notag \\
\,\Gamma &=&(19.5\pm 8.4_{-7.9}^{+5.9})~\mathrm{MeV}.  \label{eq:Data4}
\end{eqnarray}%
Parameters of $\Xi _{c}(2930)^{0}$, its mass and width were calculated in
the framework of different approaches \cite%
{Chen:2017sci,Li:2017uvv,Liu:2012sj,Wang:2017kfr,Chen:2017iyi,Ye:2017dra,Ye:2017yvl,Aliev:2018ube}%
.

The new resonances have masses and widths which do not differ considerably
from ones of $\Xi _{c}(2930)^{0}$. For simplicity of presentation, we label
parameters of $\Xi _{c}(2923)^{0}$, $\Xi _{c}(2939)^{0}$, and $\Xi
_{c}(2965)^{0}$ by subscripts $1$, $2$, and $3$, respectively. The masses
and widths of these states are equal to \cite{Aaij:2020yyt}%
\begin{eqnarray}
m_{1} &=&(2923.04\pm 0.25\pm 0.20\pm 0.14)~\mathrm{MeV},  \notag \\
\,\Gamma _{1} &=&(\,7.1\pm 0.8\pm 1.8)~\mathrm{MeV},  \label{eq:Data5}
\end{eqnarray}%
\begin{eqnarray}
m_{2} &=&(2938.55\pm 0.21\pm 0.17\pm 0.14)~\mathrm{MeV},  \notag \\
\,\Gamma _{2} &=&(\,10.2\pm 0.8\pm 1.1)~\mathrm{MeV},  \label{eq:Data6}
\end{eqnarray}%
and%
\begin{eqnarray}
m_{3} &=&(2964.88\pm 0.26\pm 0.14\pm 0.14)~\mathrm{MeV},  \notag \\
\,\Gamma _{3} &=&(\,14.1\pm 0.9\pm 1.3)~\mathrm{MeV}.  \label{eq:Data7}
\end{eqnarray}

These resonances immediately became object of theoretical investigations
\cite{Yang:2020zjl,Wang:2020gkn,Lu:2020ivo,Zhu:2020jke}, in which they were
studied in a rather detailed form. These states were considered mostly as
conventional flavor-sextet $1P$-wave baryons of different spins \cite%
{Yang:2020zjl,Wang:2020gkn} though sextet $2S$ interpretation of the
heaviest resonance from this list is also on agenda \cite{Lu:2020ivo}. The
particles $\Xi _{c}^{0}$ were described also as molecular $\overline{D}%
\Lambda -\overline{D}\Sigma $ states \cite{Zhu:2020jke}. The mass and width
of the excited flavor-antitriplet baryons $\Xi _{c}^{0}$ were calculated
recently in Ref.\ \cite{Aliev:2018ube}. Performed analysis allowed the
authors to conclude that the baryon $\Xi _{c}^{0}(1/2^{-})$ with parameters $%
\widetilde{m}=(2922\pm 83)~\mathrm{MeV}$ and $\widetilde{\Gamma }=(19.4\pm
3.3)~\mathrm{MeV}$, and quantum numbers $(1P,1/2^{-})$ may be interpreted as
the state $\Xi _{c}(2930)^{0}$. The remaining radially excited antitriplet
baryon $\Xi _{c}^{0}(1/2^{+})$ with $m^{\prime }=(2922\pm 83)~\mathrm{MeV}$
and $\Gamma ^{\prime }=(13.6\pm 2.3)~\mathrm{MeV}$ can be examined as a
candidate to one of three new resonances.

As is seen, various suggestions were made on structures and quantum numbers
of the $csd$ states, and predictions obtained by means of different methods
in the context of these assumptions, sometimes, contradict to each another.
Therefore, additional studies of these baryons are required to clarify
situation with $csd$ resonances. Before detailed analysis, there is a
necessity to establish short-hand notations for different baryons to be
studied in this article. First of all, we omit superscript $0$ for all
baryons. For the flavor-antitriplet spin-$1/2$ baryons, we use standard
notations $\Xi _{c}$, $\Xi _{c}(1/2^{-})$, and $\Xi _{c}(1/2^{+})$ for the
ground-state, first orbitally and radially excited states, respectively. The
flavor-sextet spin-$1/2$ baryons will be presented as $\Xi _{c}^{\prime }$, $%
\Xi _{c}^{\prime }(1/2^{-})$, and $\Xi _{c}^{\prime }(1/2^{+})$ in
accordance with their quantum numbers $(1S,1/2^{+})$, $(1P,1/2^{-})$, and $%
(2S,1/2^{+})$. For the spin-$3/2$ baryons with $(1S,3/2^{+})$, $(1P,3/2^{-})$%
, and $(2S,3/2^{+})$, we introduce brief notations $\Xi _{c}^{^{\ast }}$, $%
\Xi _{c}^{\ast }(3/2^{-})$, and $\Xi _{c}^{^{\ast }}(3/2^{+})$, which cannot
lead to confusions.

In the present article, we explore the ground-state $\Xi _{c}^{\prime }$ and
excited spin-$1/2$ flavor-sextet baryons $\Xi _{c}^{\prime }(1/2^{-})$ and $%
\Xi _{c}^{\prime }(1/2^{+})$ , and spin-$3/2$ particles $\Xi _{c}^{^{\ast }}$%
, $\Xi _{c}^{\ast }(3/2^{-})$, and $\Xi _{c}^{^{\ast }}(3/2^{+})$ by
computing their masses and pole residues. These spectroscopic parameters are
evaluated using the QCD sum rule method \cite{Shifman:1978bx,Shifman:1978by}%
, in which contributions of vacuum condensates up to dimension $10$ are
taken into account. We determine also widths of excited baryons $\Xi
_{c}^{\prime }(1/2^{-})$, $\Xi _{c}^{\prime }(1/2^{+})$, $\Xi _{c}^{\ast
}(3/2^{-})$, and $\Xi _{c}^{^{\ast }}(3/2^{+})$ by calculating partial
widths of their strong decays to final states $\Lambda _{c}^{+}K^{-}$ and $%
\Xi _{c}^{\prime }\pi $. Obtained predictions for partial widths of these
decay modes will allow us to estimate full widths of $\Xi _{c}^{\prime
}(1/2^{-})$, $\Xi _{c}^{\prime }(1/2^{+})$, $\Xi _{c}^{\ast }(3/2^{-})$, and
$\Xi _{c}^{^{\ast }}(3/2^{+})$. The strong decay processes are explored by
means of the QCD light-cone sum rule (LCSR) approach \cite{Balitsky:1989ry}.

This article is structured in the following way: In Sec.\ \ref{sec:MP}, we
calculate the spectroscopic parameters of the ground state and excited
baryons $\Xi _{c}^{\prime }$, $\Xi _{c}^{\prime }(1/2^{-})$, and $\Xi
_{c}^{\prime }(1/2^{+})$. Here, we also evaluate the parameters of the
states $\Xi _{c}^{\ast }$, $\Xi _{c}^{\ast }(3/2^{-})$, and $\Xi
_{c}^{^{\ast }}(3/2^{+})$. Results extracted from the sum rules in this
section will be compared with the experimental data, but also serve as input
information for the next sections. In Sec.\ \ref{sec:Decays1/2}, we derive
the LCSRs for the strong couplings $g_{1}$\ and $g_{2}$\ describing the
vertices $\Xi _{c}^{\prime }(1/2^{-})\Lambda _{c}^{+}K^{-}$\ and $\Xi
_{c}^{\prime }(1/2^{+})\Lambda _{c}^{+}K^{-}$, that are key ingredients to
evaluate width of the processes $\Xi _{c}^{\prime }(1/2^{-})\rightarrow
\Lambda _{c}^{+}K^{-}$\ and $\Xi _{c}^{\prime }(1/2^{+})\rightarrow \Lambda
_{c}^{+}K^{-}$. In Sec.\ \ref{sec:Decays1/2Pi}, we analyze the vertices $\Xi
_{c}^{\prime }(1/2^{-})\Xi _{c}^{\prime }\pi $\ and $\Xi _{c}^{\prime
}(1/2^{+})\Xi _{c}^{\prime }\pi $, and calculate corresponding strong
couplings $g_{3}$\ and $g_{4}$. The partial widths of decays $\Xi
_{c}^{\prime }(1/2^{-})\rightarrow \Xi _{c}^{\prime }\pi $\ and $\Xi
_{c}^{\prime }(1/2^{+})\rightarrow \Xi _{c}^{\prime }\pi $\ are also found
in this section. Section \ref{sec:Decays3/2} is devoted to investigation of
the decays $\Xi _{c}^{\ast }(3/2^{-})\rightarrow \Lambda _{c}^{+}K^{-},\ \Xi
_{c}^{\prime }\pi $\ and $\Xi _{c}^{^{\ast }}(3/2^{+})\rightarrow \Lambda
_{c}^{+}K^{-},\Xi _{c}^{\prime }\pi $. The last Section \ref{sec:Concl} is
reserved for comparison of obtained theoretical predictions with the LHCb
data and, in accordance with this analysis, assignment of appropriate
quantum numbers to new three LHCb resonances. This section contains also our
concluding notes. Appendix contains explicit expressions some of invariant
amplitudes employed to extract parameters of color-sextet spin-$1/2$ baryons.


\section{Masses and pole residues of the baryons $\Xi _{c}^{\prime }$ and $%
\Xi _{c}^{\ast }$}

\label{sec:MP}

The sum rules required to evaluate the mass and residue of the spin-$1/2$
baryons $\Xi _{c}^{\prime }$, $\Xi _{c}^{\prime }(1/2^{-})$, and $\Xi
_{c}^{\prime }(1/2^{+})$, and spin-$3/2$ baryons $\Xi _{c}^{^{\ast }}$, $%
\Xi _{c}^{\ast }(3/2^{-})$, and $\Xi _{c}^{^{\ast }}(3/2^{+})$ can be
obtained from analysis of the following two-point correlation functions
\begin{equation}
\Pi _{(\mu \nu )}(p)=i\int d^{4}xe^{ipx}\langle 0|\mathcal{T}\{\eta _{(\mu
)}(x)\overline{\eta }_{(\nu )}(0)\}|0\rangle ,  \label{eq:CorrF1}
\end{equation}%
where $\eta (x)$ and $\eta _{\mu }(x)$ are interpolating fields for $\Xi
_{c}^{\prime }$ and $\Xi _{c}^{\ast }$ states with spins $1/2$ and $3/2$,
respectively. In the case of the flavor-sextet spin-$1/2$ baryons the
current $\eta $ is given by the formula
\begin{eqnarray}
\eta &=&-\frac{1}{\sqrt{2}}\epsilon ^{abc}\left\{ \left(
d_{a}^{T}Cc_{b}\right) \gamma _{5}s_{c}+\beta \left( d_{a}^{T}C\gamma
_{5}c_{b}\right) s_{c}\right.  \notag \\
&&\left. -\left[ \left( c_{a}^{T}Cs_{b}\right) \gamma _{5}d_{c}+\beta \left(
c_{a}^{T}C\gamma _{5}s_{b}\right) d_{c}\right] \right\} .  \label{eq:BayC1/2}
\end{eqnarray}%
For spin-$3/2$ baryons, we use
\begin{eqnarray}
\eta _{\mu } &=&\sqrt{\frac{2}{3}}\epsilon ^{abc}\left\{ \left(
d_{a}^{T}C\gamma _{\mu }s_{b}\right) c_{c}+\left( s_{a}^{T}C\gamma _{\mu
}c_{b}\right) d_{c}\right.  \notag \\
&&\left. +\left( c_{a}^{T}C\gamma _{\mu }d_{b}\right) s_{c}\right\} .
\label{eq:BayC3/2}
\end{eqnarray}%
In formulas for the currents $C$ is the charge conjugation matrix. The
current $\eta (x)$ for the spin-$1/2$ baryons depends on an arbitrary mixing
parameter $\beta $ with $\beta =-1$ corresponding to the Ioffe current.

We begin from the spin-$1/2$ baryons and first compute the mass of the
ground-state particle $\Xi _{c}^{\prime }$. For this purposes, we express
the correlation function $\Pi ^{\mathrm{Phys}}(p)$ using the physical
parameters of the ground-state particle $(1S,\,1/2^{+})$. Then, in the
"ground-state+continuum" approximation $\Pi ^{\mathrm{Phys}}(p)$ is given by
the simple formula
\begin{equation}
\Pi ^{\mathrm{Phys}}(p)=\frac{\langle 0|\eta |\Xi _{c}^{\prime }(p,s)\rangle
\langle \Xi _{c}^{\prime }(p,s)|\overline{\eta }|0\rangle }{m^{2}-p^{2}}%
+\cdots ,  \label{eq:GS1/2}
\end{equation}%
where $m$ and $s$ are the mass and spin of $\Xi _{c}^{\prime }$,
respectively. Contributions of higher resonances and continuum states are
denoted in Eq.\ (\ref{eq:GS1/2})\ by dots. In expression for $\Pi ^{\mathrm{%
Phys}}(p)$ summation over the spin $s$ is implied.

We continue our analysis by using the matrix element
\begin{equation}
\langle 0|\eta |\Xi _{c}^{\prime }(p,s)\rangle =\lambda u(p,s),
\label{eq:MElemA}
\end{equation}%
where $\lambda $ is the pole residue of $\Xi _{c}^{\prime }$. Carrying out
 summation over $s$ in Eq.\ (\ref{eq:GS1/2}) by employing the matrix element (%
\ref{eq:MElemA}) and the formula
\begin{equation}
\sum\limits_{s}u(p,s)\overline{u}(p,s)=\slashed p+m,  \label{eq:MElemB}
\end{equation}%
we get%
\begin{equation}
\Pi ^{\mathrm{Phys}}(p)=\frac{\lambda ^{2}(\slashed p+m)}{m^{2}-p^{2}}%
+\cdots .  \label{eq:CorrF1A}
\end{equation}%
The function $\Pi ^{\mathrm{Phys}}(p)$ contains Lorentz structures
proportional to $\slashed p$ and $I$. To find the sum rule, we can employ
invariant amplitudes that correspond to these structures.

The second component of our investigation is the QCD side of the sum rule $%
\Pi ^{\mathrm{OPE}}(p)$, which should be computed in the operator product
expansion ($\mathrm{OPE}$) with certain accuracy. To this end, one has to
insert the interpolating current $\eta $ into Eq.\ (\ref{eq:CorrF1}) and
contract the quark fields. We compute $\Pi ^{\mathrm{OPE}}(p)$ using light $%
q $ and heavy $Q$ quark $x$-space propagators, explicit expressions of which
are presented below
\begin{eqnarray}
&&S_{q}^{ab}(x)=i\frac{\slashed x\delta _{ab}}{2\pi ^{2}x^{4}}-\frac{%
m_{q}\delta _{ab}}{4\pi ^{2}x^{2}}-\frac{\langle \overline{q}q\rangle \delta
_{ab}}{12}\left( 1-i\frac{m_{q}}{4}\slashed x\right)  \notag \\
&&-\frac{x^{2}\delta _{ab}}{192}\langle \overline{q}g_{s}\sigma Gq\rangle
\left( 1-i\frac{m_{q}}{6}\slashed x\right) -\frac{ig_{s}G_{ab}^{\mu \nu }}{%
32\pi ^{2}x^{2}}\left[ \slashed x\sigma _{\mu \nu }+\sigma _{\mu \nu }%
\slashed x\right]  \notag \\
&&-\frac{\slashed xx^{2}g_{s}^{2}}{7776}\langle \overline{q}q\rangle
^{2}\delta _{ab}-\frac{x^{4}\langle \overline{q}q\rangle \langle
g_{s}^{2}G^{2}\rangle }{27648}\delta _{ab}  \notag \\
&&+\frac{m_{q}g_{s}}{32\pi ^{2}}G_{ab}^{\mu \nu }\sigma _{\mu \nu }\left[
\ln \left( \frac{-x^{2}\Lambda ^{2}}{4}\right) +2\gamma _{E}\right] +\cdots ,
\label{eq:LQProp}
\end{eqnarray}%
and
\begin{eqnarray}
&&S_{Q}^{ab}(x)=\frac{m_{Q}^{2}\delta _{ab}}{4\pi ^{2}}\left[ \frac{%
K_{1}\left( m_{Q}\sqrt{-x^{2}}\right) }{\sqrt{-x^{2}}}+i\frac{\slashed %
xK_{2}\left( m_{Q}\sqrt{-x^{2}}\right) }{\left( \sqrt{-x^{2}}\right) ^{2}}%
\right]  \notag \\
&&-\frac{g_{s}m_{Q}}{16\pi ^{2}}\int_{0}^{1}duG_{ab}^{\mu \nu }(ux)\left. %
\Bigg \lbrace(\sigma _{\mu \nu }\slashed x+\slashed x\sigma _{\mu \nu
})\right.  \notag \\
&&\left. \times \frac{K_{1}\left( m_{Q}\sqrt{-x^{2}}\right) }{\sqrt{-x^{2}}}%
+2\sigma _{\mu \nu }K_{0}\left( m_{Q}\sqrt{-x^{2}}\right) \right\} .
\label{eq:HQProp}
\end{eqnarray}%
Here, $q=u,~d$ or $s$, $\gamma _{E}\simeq 0.577$ is the Euler constant, and $%
\Lambda $ is the QCD scale parameter. We also introduce the notations $%
G_{ab}^{\mu \nu }\equiv G_{A}^{\mu \nu }t_{ab}^{A}$, $G^{2}=G^{\alpha \beta
}G_{\alpha \beta }$, $A=1,2,\cdots ,8$, and $t^{A}=\lambda ^{A}/2$, with $%
\lambda ^{A}$ being the Gell-Mann matrices. The first two terms in Eq.\ (\ref%
{eq:HQProp}) in square brackets are the free part of the heavy quark
propagator in the coordinate representation, and $K_{n}(z)$ are the modified
Bessel functions of the second kind.

After performing required calculations, for $\Pi ^{\mathrm{OPE}}(p)$ we get%
\begin{equation}
\Pi ^{\mathrm{OPE}}(p)=\slashed p\Pi _{1}^{\mathrm{OPE}}(p^{2})+I\Pi _{2}^{%
\mathrm{OPE}}(p^{2}).  \label{eq:OPE}
\end{equation}%
The function $\Pi ^{\mathrm{OPE}}(p)$ expressed in terms of quark-gluon
degrees of freedom has the same Lorentz structure as $\Pi ^{\mathrm{Phys}%
}(p) $. By equating two representations of the correlation function,
performing the Borel transformation and subtracting contributions due to
higher resonances and continuum states, we extract sum rule equalities.

It is not difficult to see that the Borel transformation of $\Pi ^{\mathrm{%
Phys}}(p)$ is equal to
\begin{equation}
\mathcal{B}\Pi ^{\mathrm{Phys}}(p)=\lambda ^{2}e^{-\frac{m^{2}}{M^{2}}}(%
\slashed p+m).  \label{eq:B1}
\end{equation}%
To derive the sum rule for $m$, it is enough to use the equality
\begin{equation}
\lambda ^{2}e^{-\frac{m^{2}}{M^{2}}}=\Pi _{1}^{\mathrm{OPE}}(M^{2},s_{0}),\
\label{eq:SRequality}
\end{equation}%
where $\Pi _{1}^{\mathrm{OPE}}(M^{2},s_{0})$ is the Borel transformed and
subtracted invariant amplitude $\Pi _{1}^{\mathrm{OPE}}(p^{2})$, and $M^{2}$
and $s_{0}$ are the Borel and continuum threshold parameters, respectively.
The sum rule for the mass of the ground state particle has a simple form and
can be found from the formula
\begin{equation}
m^{2}=\frac{\Pi _{1}^{\prime \mathrm{OPE}}(M^{2},s_{0})}{\Pi _{1}^{\mathrm{%
OPE}}(M^{2},s_{0})}.  \label{eq:SR1A}
\end{equation}%
In Eq.\ (\ref{eq:SR1A}), we use the short-hand notation $\Pi _{1}^{\prime
\mathrm{OPE}}(M^{2},s_{0})=d/d(-1/M^{2})\Pi _{1}^{\mathrm{OPE}}(M^{2},s_{0})$%
.

In order to derive sum rules for parameters of the excited states, let us
note that the current $\eta $ couples not only to ground-state particle $\Xi
_{c}^{\prime }$, but also to its first orbital and radial excitations $\Xi
_{c}^{\prime }(1/2^{-})$, and $\Xi _{c}^{\prime }(1/2^{+})$, respectively.
To write down the phenomenological side of the sum rule, we use the
"ground-state+excited-state+continuum" scheme. Therefore, we take into
account effects of the baryons $\Xi _{c}^{\prime }$ and $\Xi _{c}^{\prime
}(1/2^{-})$, and find
\begin{eqnarray}
&&\Pi ^{\mathrm{Phys}}(p)=\frac{\langle 0|\eta |\Xi _{c}^{\prime
}(p,s)\rangle \langle \Xi _{c}^{\prime }(p,s)|\overline{\eta }|0\rangle }{%
m^{2}-p^{2}}  \notag \\
&&+\frac{\langle 0|\eta |\Xi _{c}^{\prime }(1/2^{-},p,\widetilde{s})\rangle
\langle \Xi _{c}^{\prime }(1/2^{-},p,\widetilde{s})|\overline{\eta }%
||0\rangle }{\widetilde{m}^{2}-p^{2}}+\cdots ,  \notag \\
&&  \label{eqCF1/2}
\end{eqnarray}%
where $\widetilde{m}$ and $\widetilde{s}$ are the mass and spin of the
excited state $\Xi _{c}^{\prime }(1/2^{-})$.

To simplify $\Pi ^{\mathrm{Phys}}(p)$, we employ Eq.\ (\ref{eq:MElemA}) and
additionally introduce the matrix element
\begin{equation}
\langle 0|\eta |\Xi _{c}^{\prime }(1/2^{-},p,\widetilde{s})\rangle =%
\widetilde{\lambda }\gamma _{5}\widetilde{u}(p,\widetilde{s}),
\label{eq:MElem}
\end{equation}%
where $\widetilde{\lambda }$ is the pole residue of the baryon $\Xi
_{c}^{\prime }(1/2^{-})$. Performing  summations
over $s$ and $\widetilde{s}$ in Eq.\ (\ref{eqCF1/2}) by employing relevant matrix elements and the
formula (\ref{eq:MElemB}), we get%
\begin{equation}
\Pi ^{\mathrm{Phys}}(p)=\frac{\lambda ^{2}(\slashed p+m)}{m^{2}-p^{2}}+\frac{%
\widetilde{\lambda }^{2}(\slashed p-\widetilde{m})}{\widetilde{m}^{2}-p^{2}}%
+\cdots .  \label{eq:CorrF2}
\end{equation}%
The Borel transformation of $\Pi ^{\mathrm{Phys}}(p)$ is equal to
\begin{eqnarray}
&&\mathcal{B}\Pi ^{\mathrm{Phys}}(p)=\lambda ^{2}e^{-\frac{m^{2}}{M^{2}}}(%
\slashed p+m)  \notag \\
&&+\widetilde{\lambda }^{2}e^{-\frac{\widetilde{m}^{2}}{M^{2}}}(\slashed p-%
\widetilde{m}).  \label{eq:Bor1}
\end{eqnarray}%
Then, the sum rule equalities are
\begin{equation}
\lambda ^{2}e^{-\frac{m^{2}}{M^{2}}}+\widetilde{\lambda }^{2}e^{-\frac{%
\widetilde{m}^{2}}{M^{2}}}=\Pi _{1}^{\mathrm{OPE}}(M^{2},\widetilde{s}_{0}),
\label{eq:MFor1}
\end{equation}%
and
\begin{equation}
\lambda ^{2}me^{-\frac{m^{2}}{M^{2}}}-\widetilde{\lambda }^{2}\widetilde{m}%
e^{-\frac{\widetilde{m}^{2}}{M^{2}}}=\Pi _{2}^{\mathrm{OPE}}(M^{2},%
\widetilde{s}_{0}).  \label{eq:MFor2}
\end{equation}%
The first of these expressions is extracted from the structure $\sim \slashed
p$, whereas the second one corresponds to terms proportional to $I$.

The derived equalities (\ref{eq:MFor1}) and (\ref{eq:MFor2}) contain four
parameters $(m,\ \lambda )$ and $(\widetilde{m},\ \widetilde{\lambda })$ of
the ground state and first orbitally excited baryon. As the mass $m$ of the
ground-state baryon $\Xi _{c}^{\prime }$, we use its value evaluated from
the sum rule (\ref{eq:SR1A}). Therefore, one has to find sum rules for the
pole residue of the ground-state particle, as well as parameters $(%
\widetilde{m},\ \widetilde{\lambda })$ of the excited state. Usual way to
handle this problem is to act by the operator $d/d(-1/M^{2})$ to Eqs.\ (\ref%
{eq:MFor1}) and (\ref{eq:MFor2}), and get missing equations. Then, after
simple manipulations, we obtain
\begin{eqnarray}
\widetilde{m}^{2} &=&\frac{\Pi _{2}^{\prime \mathrm{OPE}}-m\Pi _{1}^{\prime
\mathrm{OPE}}}{\Pi _{2}^{\mathrm{OPE}}-m\Pi _{1}^{\mathrm{OPE}}},  \notag \\
\lambda ^{2} &=&\frac{\widetilde{m}\Pi _{1}^{\mathrm{OPE}}+\Pi _{2}^{\mathrm{%
OPE}}}{m+\widetilde{m}}e^{m^{2}/M^{2}}  \notag \\
\widetilde{\lambda }^{2} &=&\frac{m\Pi _{1}^{\mathrm{OPE}}-\Pi _{2}^{\mathrm{%
OPE}}}{m+\widetilde{m}}e^{\widetilde{m}^{2}/M^{2}}.  \label{eq:SR1}
\end{eqnarray}%
Expressions written down in Eq.\ (\ref{eq:SR1}) are the QCD two-point sum
rules for parameters of the ground-state and excited baryon, which can be
employed to evaluate their numerical values. In these formulas, for
simplicity, we do not show dependence of the functions $\Pi _{1,2}^{(\prime )%
\mathrm{OPE}}(M^{2},\widetilde{s}_{0})$ on $M^{2}$ and $\widetilde{s}_{0}$.
The parameters of the radially excited baryon $\Xi _{c}^{\prime }(1/2^{+})$
can be extracted using $\Pi ^{\mathrm{Phys}}(p)$, in which the excited $1P$
state is replaced by $2S$ particle. In Eq.\ (\ref{eq:SR1}) this is
equivalent to transformation $\widetilde{m}\rightarrow -m^{\prime }$, and
redefinition of the residue $\widetilde{\lambda }\rightarrow \lambda
^{\prime }$, where $(m^{\prime }$, $\lambda ^{\prime })$ are parameters of $%
\Xi _{c}^{\prime }(1/2^{+})$.

The sum rules (\ref{eq:SR1A}) and (\ref{eq:SR1}) depend on the vacuum
expectations values of the different quark, gluon, and mixed operators, as
well as on the masses of $s$ and $c$-quarks. Values of these universal input
parameters are presented below
\begin{eqnarray}
&&\langle \overline{q}q\rangle =-(0.24\pm 0.01)^{3}~\mathrm{GeV}^{3},\
\langle \overline{s}s\rangle =0.8\langle \bar{q}q\rangle ,  \notag \\
&&\langle \overline{q}g_{s}\sigma Gq\rangle =m_{0}^{2}\langle \overline{q}%
q\rangle ,\ \langle \overline{s}g_{s}\sigma Gs\rangle =m_{0}^{2}\langle
\overline{s}s\rangle,  \notag \\
&&m_{0}^{2}=(0.8\pm 0.1)~\mathrm{GeV}^{2},  \notag \\
&&\langle \frac{\alpha _{s}G^{2}}{\pi }\rangle =(0.012\pm 0.004)~\mathrm{GeV}%
^{4},  \notag \\
&&m_{s}=93_{-5}^{+11}~\mathrm{MeV},\ m_{c}=1.27\pm 0.2~\mathrm{GeV}.
\label{eq:Parameters}
\end{eqnarray}

The sum rules contain also auxiliary parameters $M^{2}$ and $s_{0}$, which
are not arbitrary, but should meet some restrictions. Thus, inside of
working regions of these parameters convergence of the operator product
expansion should be fulfilled. The dominance of the pole contribution, and
prevalence of the perturbative term in the sum rules are also among
constraints of computations. The extracted predictions should be stable
against variations of $M^{2}$ and $\beta $: the latter is necessary for spin-%
$1/2$ particles. In order to explore the dependence on $\beta $, it is
convenient to introduce a parameter $\cos \theta $ through $\beta =\tan
\theta $.

Calculations of $\Pi ^{\mathrm{OPE}}(p)$ are performed by including into
analysis nonperturbative terms till dimension $10$. In computations we set $%
m_{d}=0$, but take into account terms $\sim m_{s}$. Explicit expressions of
the amplitudes $\Pi _{1}^{\mathrm{OPE}}(M^{2},s_{0})$ and $\Pi _{2}^{\mathrm{%
OPE}}(M^{2},s_{0})$ in simple case $m_{s}=0$ can be found in Appendix.

First, we calculate the mass of the ground-state particle $\Xi _{c}^{\prime
} $. The parameters $M^{2}$ and $s_{0}$ necessary for such analysis and
prediction obtained for $m$ are presented in Table\ \ref{tab:Results1A}.
Results for the masses and residues of the baryons $\Xi _{c}^{\prime
}(1/2^{-})$ and $\Xi _{c}^{\prime }(1/2^{+})$ are also collected in Table\ %
\ref{tab:Results1A}. In this Table, we write down the working regions for $%
M^{2}$ and $\widetilde{s}_{0}$ used to evaluate $\widetilde{m},$ $\widetilde{%
\lambda }$, and $\lambda $. The auxiliary parameter $\cos \theta $ has been
varied inside of the boundaries
\begin{equation}
-1.0\leq \cos \theta \leq -0.5,\ 0.5\leq \cos \theta \leq 1.0,
\label{eq:Beta}
\end{equation}%
where we have attained best stability for our predictions. In numerical
analysis, we use average values of quantities calculated at $\cos \theta
=\pm 0.75$ from these regions.

\begin{table}[tbp]
\begin{tabular}{|c|c|c|c|}
\hline\hline
Baryons & $\Xi _{c}^{\prime}$ & $\Xi _{c}^{\prime }(1/2^{-})$ & $\Xi
_{c}^{\prime }(1/2^{+})$ \\ \hline
$(n, J^{\mathrm{P}})$ & $(1S, \frac{1}{2}^{+})$ & $(1P, \frac{1}{2}^{-})$ & $%
(2S, \frac{1}{2}^{+})$ \\
$M^2 ~(\mathrm{GeV}^2$) & $3-5$ & $3-5$ & $3-5$ \\
$s_0 ~(\mathrm{GeV}^2$) & $2.8^2-3.0^2$ & $3.2^2-3.4^2$ & $3.2^2-3.4^2$ \\
$m ~(\mathrm{MeV})$ & $2576 \pm 150$ & $2925 \pm 115$ & $2925 \pm 115$ \\
$\lambda\times 10^{2} ~(\mathrm{GeV}^3)$ & $4.0 \pm 0.5$ & $3.9 \pm 1.3$ & $%
15.4 \pm 5.0$ \\ \hline\hline
\end{tabular}%
\caption{The sum rule results for the masses and residues of the spin-$1/2$
flavor-sextet baryons $\Xi _{c}^{\prime}$.}
\label{tab:Results1A}
\end{table}

\begin{figure}[h]
\begin{center}
\includegraphics[totalheight=6cm,width=8cm]{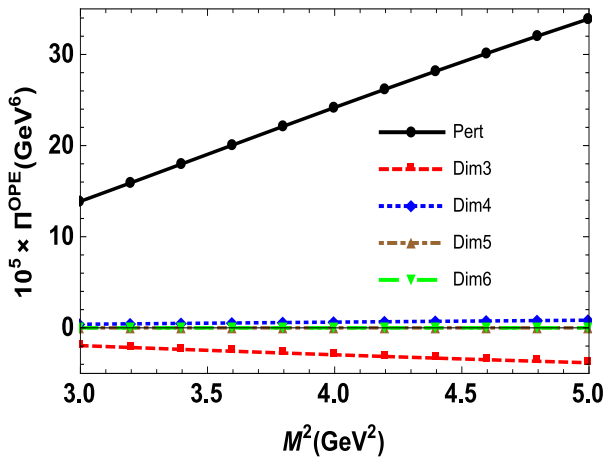}
\end{center}
\caption{ The different contributions to $\Pi_{1} ^{\mathrm{OPE}}(M^2,s_0)$
at fixed $s_{0}=3.3^{2}~\mathrm{MeV}^{2}$.}
\label{fig:CF}
\end{figure}

The amplitudes $\Pi _{1(2)}^{\mathrm{OPE}}(M^{2},s_{0})$ are formed mainly
due to perturbative contributions. In Fig.\ \ref{fig:CF}, as an example, we
present different components of the amplitude $\Pi _{1}^{\mathrm{OPE}%
}(M^{2},s_{0})$ as functions of $M^{2}$ at fixed $s_{0}=3.3^{2}~\mathrm{MeV}%
^{2}$. It is seen, that perturbative term is a dominant contribution to $\Pi
_{1}^{\mathrm{OPE}}$. Nonperturbative effects are small, and only $\mathrm{%
Dim3}$ term with quark condensate factor $\langle \overline{s}s\rangle
+\langle \overline{d}d\rangle $ can be considered as essential one. Higher
dimensional contributions $N>6$ are very small: Corresponding curves are
undistinguishable in the plot, for this reason, we do not included them into
the figure. In the case under consideration, the operator product expansion
demonstrates rapid convergence. Thus, at $M^{2}=4~\mathrm{MeV}^{2}$ already $%
\mathrm{Dim5}$ term is less than $0.01$ of the perturbative contribution.

Reliable predictions for physical quantities imply dominance of the pole
contribution ($\mathrm{PC}$) in the sum rule analysis. In the
"ground-state+continuum" scheme fixed by $s_{0}\sim 2.9^{2}\ ~\mathrm{MeV}%
^{2}$, a $\mathrm{PC}$ higher than $50\%$ of the whole result leads to
credible predictions for parameters of the ground-state baryon. In the
"ground-state +first excited state+continuum" scheme determined by $%
\widetilde{s}_{0}\sim 3.3^{2}\ ~\mathrm{MeV}^{2}$, there are two particles
that generate the pole contribution. In our case, these two baryons
constitute $79\%$ of the total contribution in average. In other words,
excited state form approximately $25\%$ of the whole contribution. This is
less than $50\%$ limit necessary for the ground-state or isolated excited
particle. But mass and pole residue of the excited baryon are obtained from
expressions, which contain contributions of the ground-state baryon as well.
For these expressions, as we have noted above, $\mathrm{PC}\approx 79\%$
which assures correctness of extracted quantities. It is worth noting that
continuum threshold parameters $s_{0}$ and $\widetilde{s}_{0}$ in this two
schemes obey the restriction $s_{0}<\widetilde{s}_{0}$ which demonstrates
the self-consistency of performed analyses. The central values of the masses
of the excited baryons $\widetilde{m}$ $=m^{\prime }$ are above $\sqrt{s_{0}}
$ and below $\sqrt{\widetilde{s}_{0}}$, as they should be.

\begin{figure}[h]
\begin{center}
\includegraphics[totalheight=6cm,width=8cm]{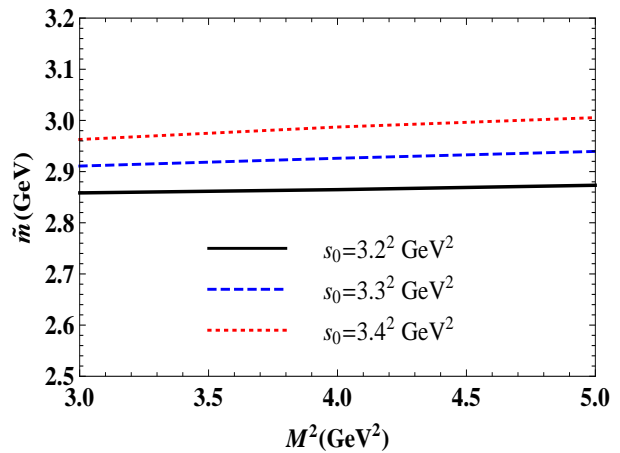}
\end{center}
\caption{ The mass of the baryon $\Xi _{c}^{\prime }(1/2^{-})$ as a function
of the parameter $M^{2}$ at fixed $s_{0}$.}
\label{fig:Mass1}
\end{figure}

\begin{widetext}

\begin{figure}[h!]
\begin{center}
\includegraphics[totalheight=6cm,width=8cm] {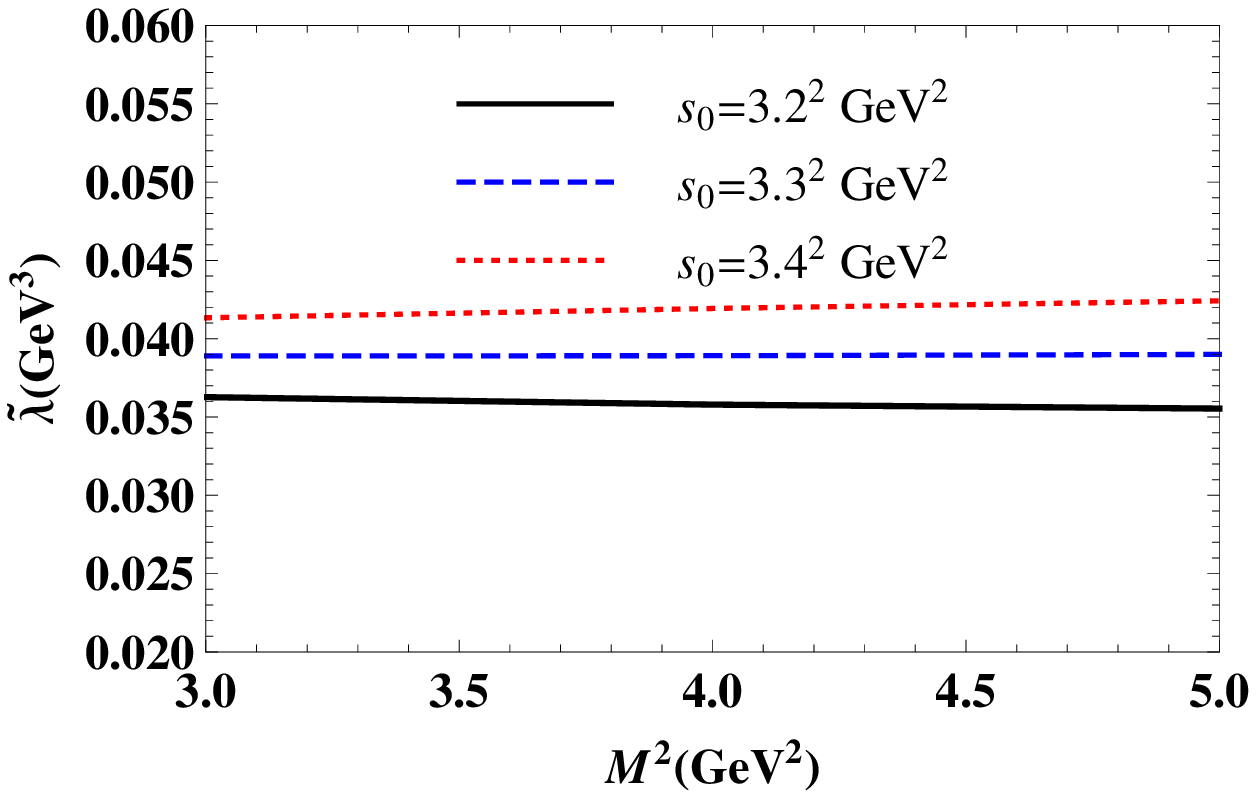}%
\includegraphics[totalheight=6cm,width=8cm]{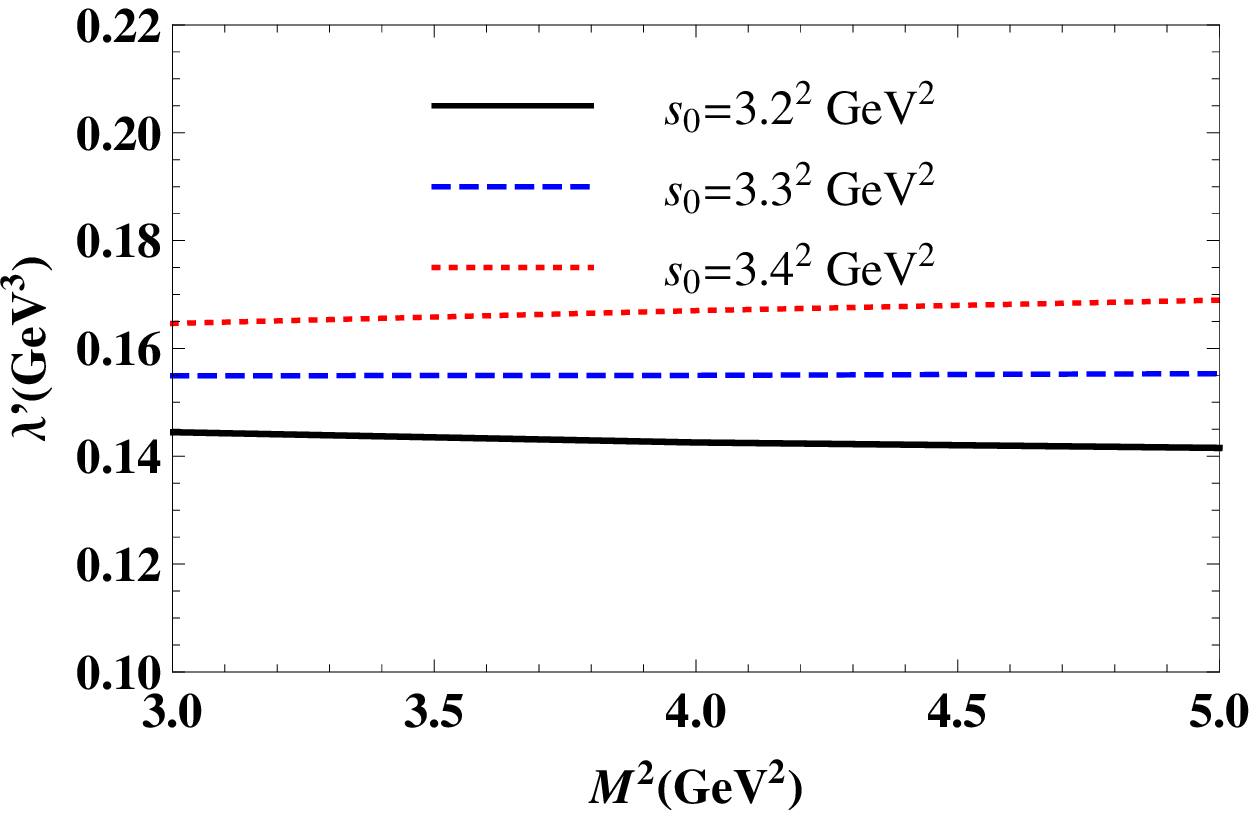}
\end{center}
\caption{ The dependence of the residues $\widetilde{\lambda}$ (left panel) and $\lambda^{\prime}$ (right panel) on the Borel parameter $M^2$ and at fixed $s_0$.  }
\label{fig:Residue1}
\end{figure}

\end{widetext}

In Fig.\ \ref{fig:Mass1}, we plot the mass of the particle $\Xi _{c}^{\prime
}(1/2^{-})$ as a function of $M^{2}$. Here, one can see dependence of the
obtained result on the Borel parameter $M^{2}$, which have been pictured at
fixed values of the continuum threshold parameter $s_{0}$. The residues of
the excited baryons $\Xi _{c}^{\prime }(1/2^{-})$ and $\Xi _{c}^{\prime
}(1/2^{+})$ are drawn in Fig.\ \ref{fig:Residue1}, where sensitivity of $\
\widetilde{\lambda }$ and $\lambda ^{\prime }$ to a choice of $M^{2}$ is
shown. All these parameters are very stable against variation of the Borel
parameter. The main part of theoretical uncertainties come from variation of
the continuum threshold parameter $s_{0}$, which is also seen in these
figures.

The similar analysis with some new technical details can be carried out for
the spin-$3/2$ baryons $\Xi _{c}^{\ast }$ as well. Indeed, in this case, in
order to find the physical side of the sum rules, we use the matrix elements
\begin{eqnarray}
&&\langle 0|\eta _{\mu }|\Xi _{c}^{\ast }(p,s)\rangle =\lambda ^{\ast
}u_{\mu }(p,s),  \notag \\
&&\langle 0|\eta _{\mu }|\Xi _{c}^{\ast }(3/2^{-},p,\widetilde{s})\rangle =%
\widetilde{\lambda }^{\ast }\gamma _{5}\widetilde{u}_{\mu }(p,\widetilde{s}),
\label{eq:MElem2}
\end{eqnarray}%
where $u_{\mu }(p,s)$ and $\widetilde{u}_{\mu }(p,\widetilde{s})$ are the
Rarita-Schwinger spinors, and perform the summation over the spins $s$ and $%
\widetilde{s}$ using the expression
\begin{equation}
\sum\limits_{s}u_{\mu }(p,s)\overline{u}_{\nu }(p,s)=-(\slashed p+m^{\ast
})F_{\mu \nu }(m^{\ast },p),  \label{eq:ME2}
\end{equation}%
where
\begin{eqnarray}
F_{\mu \nu }(m^{\ast },p) &=&\left[ g_{\mu \nu }-\frac{1}{3}\gamma _{\mu
}\gamma _{\nu }-\frac{2}{3m^{\ast 2}}p_{\mu }p_{\nu }\right.  \notag \\
&&\left. +\frac{1}{3m^{\ast }}(p_{\mu }\gamma _{\nu }-p_{\nu }\gamma _{\mu })%
\right].  \label{eq:Eff}
\end{eqnarray}%
Here, $m^{\ast }$ is the mass of the spin-$3/2$ baryons $\Xi _{c}^{\ast }$.

In calculations one should take into account that the interpolating current $%
\eta _{\mu }$ couples both to spin-$3/2$ and spin-$1/2$ baryons. Therefore,
the sum rules contain contributions of spin-$1/2$ particles as well. These
terms should be removed by using a special ordering of the Dirac matrices.
Indeed, it is easy to demonstrate that structures $\sim \slashed pg_{\mu \nu
}$ and $\sim g_{\mu \nu }$ are formed only due to contributions of spin-$3/2$
baryons. Therefore, to find the sum rules for the mass and residue of the
ground state particle $\Xi _{c}^{\ast }$, and parameters of the excited
baryons $\Xi _{c}^{\ast }(3/2^{-})$, and $\Xi _{c}^{^{\ast }}(3/2^{+})$, we
use only these structures and corresponding invariant amplitudes.

To determine the QCD side of the sum rules, the correlation function $\Pi
_{\mu \nu }(p)$ has to be computed also in terms of the quark propagators.
We calculate $\Pi _{\mu \nu }^{\mathrm{OPE}}(p)$ by utilizing Eq.\ (\ref%
{eq:CorrF1}) and the current given by Eq.\ (\ref{eq:BayC3/2}). Operations to
find $\Pi _{\mu \nu }^{\mathrm{OPE}}(p)$ using the quark propagators in the $%
x$-space and calculation of the Borel transformed and subtracted invariant
amplitudes are well known and were presented in the literature (for instance
see \cite%
{Agaev:2016srl,Aliev:2011ufa,Azizi:2015ksa,Aliev:2018ube,Azizi:2018duk}).
Thus, we do not go into further details of these computations, and emphasize
only that analysis has been performed with dimension-$10$ accuracy. We also
note that the final results are very lengthy, so we do not present their
explicit expressions here.

Results obtained for parameters of the spin-$3/2$ baryons $\Xi _{c}^{\ast }$,%
$\ \Xi _{c}^{\ast }(3/2^{-})$, and $\Xi _{c}^{^{\ast }}(3/2^{+})$ are
presented in Table\ \ref{tab:Results2A}. Here, we write down the working
regions for parameters $M^{2}$ and $s_{0}$ used to evaluate $m^{\ast }$ and $%
\lambda ^{\ast }$.

\begin{table}[tbp]
\begin{tabular}{|c|c|c|c|}
\hline\hline
Baryons & $\Xi_{c}^{\star}$ & $\Xi_{c}^{\star}(3/2^{-})$ & $%
\Xi_{c}^{\star}(3/2^{+})$ \\ \hline
$(n, J^{\mathrm{P}})$ & $(1S, \frac{3}{2}^{+})$ & $(1P, \frac{3}{2}^{-})$ & $%
(2S, \frac{3}{2}^{+})$ \\
$M^2 ~(\mathrm{GeV}^2$) & $3-5$ & $3-5$ & $3-5$ \\
$s_0 ~(\mathrm{GeV}^2$) & $3.0^2-3.2^2$ & $3.2^2-3.4^2$ & $3.2^2-3.4^2$ \\
$m^{*} ~(\mathrm{MeV})$ & $2655 \pm 102$ & $2960 \pm 67$ & $2960 \pm 67$ \\
$\lambda^{*}\times 10^{2} ~(\mathrm{GeV}^3)$ & $4.7 \pm 0.4$ & $2.4 \pm 0.3$
& $10.4 \pm 1.4$ \\ \hline\hline
\end{tabular}%
\caption{The predictions for spectroscopic parameters of the spin-$3/2$
baryons $\Xi _{c}^{\ast }$.}
\label{tab:Results2A}
\end{table}

The masses and residues of the baryons $\Xi _{c}^{\ast }$ as functions of
the parameters $M^{2}$ and $s_{0}$ demonstrate behavior similar to ones of
the spin-$1/2$ particles, therefore we do not provide corresponding
graphics, by noting that systematic errors of calculations do not exceed
limits accepted in the sum rule method.

As is seen, the sum rule method employed in the present work to find masses
of the spin-$1/2$ and -$3/2$ baryons $\Xi _{c}^{\prime }$ and $\Xi
_{c}^{\ast }$ leads for the first orbitally and radially excited states to
the same predictions. Therefore, relying only on this information, it is
impossible to make assignment for three new resonances observed by the LHCb
collaboration. To compare with LHCb experimental data, one needs to
determine also widths of these particles.


\section{ Decays of $\Xi _{c}^{\prime }(1/2^{-})$ and $\Xi _{c}^{\prime
}(1/2^{+})$ to $\Lambda _{c}^{+}K^{-}$}

\label{sec:Decays1/2}
In this section we study the vertices $\Xi _{c}^{\prime }(1/2^{-})\Lambda
_{c}^{+}K^{-}$ and $\Xi _{c}^{\prime }(1/2^{+})\Lambda _{c}^{+}K^{-}$, and
calculate corresponding strong couplings, which are required to compute
width of the decays $\Xi _{c}^{\prime }(1/2^{-})\rightarrow \Lambda
_{c}^{+}K^{-}$ and $\Xi _{c}^{\prime }(1/2^{+})\rightarrow \Lambda
_{c}^{+}K^{-}$, respectively. There are different sum rule methods to
extract numerical values of these couplings. They can be calculated using
both the QCD three-point and light-cone sum rule methods. But LCSR method
have some advantages compared to the three-point sum rule approach in
calculations of the strong couplings and form factors. The reason is that in
the three-point sum rules higher orders in OPE are enhanced by powers of the
heavy quark mass, and for sufficiently large masses OPE breaks down. The
LCSR method does not suffer from such problems, because it is consistent
with the heavy-quark limit and provides more effective tools for
investigations than alternative approaches \cite{Belyaev:1994zk}. Therefore,
for analyses of the vertices $\Xi _{c}^{\prime }(1/2^{-})\Lambda
_{c}^{+}K^{-}$ and $\Xi _{c}^{\prime }(1/2^{+})\Lambda _{c}^{+}K^{-}$, we
use the QCD LCSR method.

To this end, we start from analysis of the correlation function%
\begin{equation}
\Pi (p,q)=i\int d^{4}xe^{ipx}\langle K(q)|\mathcal{T}\{\eta _{\Lambda }(x)%
\overline{\eta }(0)\}|0\rangle ,  \label{eq:CorrF3}
\end{equation}%
where $\eta _{\Lambda }(x)$ is the interpolating field for the $\Lambda _{c}$
baryon. The $\Lambda _{c}$ is the flavor-antitriplet spin-$1/2$ particle,
and its current is given by the expression
\begin{eqnarray}
\eta _{\Lambda } &=&\frac{1}{\sqrt{6}}\epsilon ^{abc}\left\{ 2\left(
u_{a}^{T}Cd_{b}\right) \gamma _{5}c_{c}+2\widetilde{\beta }\left(
u_{a}^{T}C\gamma _{5}d_{b}\right) c_{c}\right.  \notag \\
&&+\left( u_{a}^{T}Cc_{b}\right) \gamma _{5}d_{c}+\widetilde{\beta }\left(
u_{a}^{T}C\gamma _{5}c_{b}\right) d_{c}  \notag \\
&&\left. +\left( c_{a}^{T}Cd_{b}\right) \gamma _{5}u_{c}+\widetilde{\beta }%
\left( c_{a}^{T}C\gamma _{5}d_{b}\right) u_{c}\right\} ,
\label{eq:LambdaCurr}
\end{eqnarray}%
where $\widetilde{\beta }$ is the arbitrary mixing parameter.

First, we write the correlation function $\Pi (p,q)$ in terms of involved
baryons' parameters, and find by this way the physical or hadronic side of
the sum rule. As a result, we obtain%
\begin{eqnarray}
&&\Pi ^{\mathrm{Phys}}(p,q)=\frac{\langle 0|\eta _{\Lambda }|\Lambda
_{c}^{+}(p,s)\rangle }{p^{2}-m_{\Lambda }^{2}}\langle K(q)\Lambda
_{c}^{+}(p,s)|\Xi _{c}^{\prime }(p^{\prime },s^{\prime })\rangle  \notag \\
&&\times \frac{\langle \Xi _{c}^{\prime }(p^{\prime },s^{\prime })|\overline{%
\eta }|0\rangle }{p^{\prime 2}-m^{2}}+\frac{\langle 0|\eta _{\Lambda
}|\Lambda _{c}^{-}(p,s)\rangle }{p^{2}-\widetilde{m}_{\Lambda }^{2}}  \notag
\\
&&\times \langle K(q)\Lambda _{c}^{-}(p,s)|\Xi _{c}^{\prime }(p^{\prime
},s^{\prime })\rangle \frac{\langle \Xi _{c}^{\prime }(p^{\prime },s^{\prime
})|\overline{\eta }|0\rangle }{p^{\prime 2}-m^{2}}  \notag \\
&&+\frac{\langle 0|\eta _{\Lambda }|\Lambda _{c}^{+}(p,s)\rangle }{%
p^{2}-m_{\Lambda }^{2}}\langle K(q)\Lambda _{c}^{+}(p,s)|\Xi _{c}^{\prime
}(1/2^{-},p^{\prime },s^{\prime })\rangle  \notag \\
&&\times \frac{\langle \Xi _{c}^{\prime }(1/2^{-},p^{\prime },s^{\prime })|%
\overline{\eta }|0\rangle }{p^{\prime 2}-\widetilde{m}^{2}}+\frac{\langle
0|\eta _{\Lambda }|\Lambda _{c}^{-}(p,s)\rangle }{p^{2}-\widetilde{m}%
_{\Lambda }^{2}}  \notag \\
&&\times \langle K(q)\Lambda _{c}^{-}(p,s)|\Xi _{c}^{\prime
}(1/2^{-},p^{\prime },s^{\prime })\rangle  \notag \\
&&\times \frac{\langle \Xi _{c}^{\prime }(1/2^{-},p^{\prime },s^{\prime })|%
\overline{\eta }|0\rangle }{p^{\prime 2}-\widetilde{m}^{2}}+\cdots ,
\label{eq:CorrF4}
\end{eqnarray}%
where $p^{\prime }=p+q,\ p$ and $q$ are the momenta of the $\Xi _{c}^{\prime
},$ $\Lambda _{c}$ baryons and $K$ meson, respectively. The $\Lambda
_{c}^{+} $ and $\Lambda _{c}^{-}$ are baryons with quantum numbers $%
(1S,1/2^{+})$ and $(1P,1/2^{-})$, and masses $m_{\Lambda }$ and $\widetilde{m%
}_{\Lambda }$, respectively. The dots in Eq.\ (\ref{eq:CorrF4}) stand for
contributions of the higher resonances and continuum states.

To continue, we introduce the matrix elements of the baryons $\Lambda _{c}$
\begin{eqnarray}
\langle 0|\eta _{\Lambda }|\Lambda _{c}^{+}(p,s)\rangle &=&\lambda _{\Lambda
}u(p,s),  \notag \\
\langle 0|\eta _{\Lambda }|\Lambda _{c}^{-}(p,s)\rangle &=&\widetilde{%
\lambda }_{\Lambda }\gamma _{5}u(p,s),  \label{eq:ME3}
\end{eqnarray}%
and also parametrize remaining unknown matrix elements in terms of the
strong couplings \cite{Aliev:2010yx,Aliev:2018ube}
\begin{eqnarray}
&&\langle K(q)\Lambda _{c}^{+}(p,s)|\Xi _{c}^{\prime }(p^{\prime },s^{\prime
})\rangle =g_{0}\overline{u}(p,s)\gamma _{5}u(p^{\prime },s^{\prime }),
\notag \\
&&\langle K(q)\Lambda _{c}^{-}(p,s)|\Xi _{c}^{\prime }(p^{\prime },s^{\prime
})\rangle =\widetilde{g}_{0}\overline{u}(p,s)u(p^{\prime },s^{\prime }),
\notag \\
&&\langle K(q)\Lambda _{c}^{+}(p,s)|\Xi _{c}^{\prime }(1/2^{-},p^{\prime
},s^{\prime })\rangle =g_{1}\overline{u}(p,s)u(p^{\prime },s^{\prime }),
\notag \\
&&{}\langle K(q)\Lambda _{c}^{-}(p,s)|\Xi _{c}^{\prime }(1/2^{-},p^{\prime
},s^{\prime })\rangle =\widetilde{g}_{1}\overline{u}(p,s)\gamma
_{5}u(p^{\prime },s^{\prime }),  \notag \\
&&  \label{eq.ME4}
\end{eqnarray}%
where $\lambda _{\Lambda }$ and $\widetilde{\lambda }_{\Lambda }$ are pole
residues of $\Lambda _{c}^{+}$ and $\Lambda _{c}^{-}$, respectively.

Then using the matrix elements of the particles $\Xi _{c}^{\prime }$ and $%
\Xi _{c}^{\prime }(1/2^{-})$, carrying out the summations over the spins $s$
and $s^{\prime }$, and applying the double Borel transformation with respect
$p^{2}$ and $p^{\prime 2}$, for the phenomenological side of the sum rules,
we obtain
\begin{eqnarray}
&&\mathcal{B}\Pi ^{\mathrm{Phys}}(p^{2},p^{\prime 2})=g_{0}\lambda \lambda
_{\Lambda }e^{-m^{2}/M_{1}^{2}}e^{-m_{\Lambda }^{2}/M_{2}^{2}}(\slashed %
p+m_{\Lambda })  \notag \\
&&\times \gamma _{5}\left( \slashed p^{\prime }+m\right) -\widetilde{g}%
_{0}\lambda \widetilde{\lambda }_{\Lambda }e^{-m^{2}/M_{1}^{2}}e^{-%
\widetilde{m}_{\Lambda }^{2}/M_{2}^{2}}(\slashed p-\widetilde{m}_{\Lambda })
\notag \\
&&\times \gamma _{5}\left( \slashed p^{\prime }+m\right) +g_{1}\widetilde{%
\lambda }\lambda _{\Lambda }e^{-\widetilde{m}^{2}/M_{1}^{2}}e^{-m_{\Lambda
}^{2}/M_{2}^{2}}(\slashed p+m_{\Lambda })  \notag \\
&&\times \gamma _{5}\left( \slashed p^{\prime }-\widetilde{m}\right) -%
\widetilde{g}_{1}\widetilde{\lambda }\widetilde{\lambda }_{\Lambda }e^{-%
\widetilde{m}^{2}/M_{1}^{2}}e^{-\widetilde{m}_{\Lambda }^{2}/M_{2}^{2}}
\notag \\
&&\times (\slashed p-\widetilde{m}_{\Lambda })\gamma _{5}\left( \slashed %
p^{\prime }-\widetilde{m}\right) ,  \label{eq:CorrF5}
\end{eqnarray}%
where $M_{1}^{2}$ and $M_{2}^{2}$ are the Borel parameters.

As is seen, Eq.\ (\ref{eq:CorrF5}) contains different Lorentz structures. To
extract sum rules, it is convenient to reorganize these terms into
structures proportional to $\slashed q\slashed p\gamma _{5}$, $\slashed %
p\gamma _{5}$, $\slashed q\gamma _{5}$ and $\gamma _{5}$, and employ
corresponding invariant amplitudes. The same structures appear in the QCD
side of the sum rule equality, which has to be calculated using the quark
propagators. After performing the double Borel transformation of $\Pi ^{%
\mathrm{OPE}}(p,q)$, we get $\mathcal{B}\Pi ^{\mathrm{OPE}}(p^{2},p^{\prime
2})=\Pi ^{\mathrm{OPE}}(M_{1}^{2},M_{2}^{2})$ which is a function of two
Borel parameters. To proceed, it is convenient to introduce $M^{2}$ through
the relation
\begin{equation}
\frac{1}{M^{2}}=\frac{1}{M_{1}^{2}}+\frac{1}{M_{2}^{2}},  \label{eq:Trick}
\end{equation}%
and use $M_{1}^{2}=M_{2}^{2}=2M^2 $ to go from the double-dispersion
integral to the single integral representation by performing one of the
dispersion integrals. We set $M_{1}^{2}=M_{2}^{2} $ as the masses of the $%
\Xi _{c}$ and $\Lambda _{c}$ baryons are close, and uncertainties expected
due to this choice are small. As a result, we get a single integral
representation for $\Pi ^{\mathrm{OPE}}\left( M^{2}\right) $, which
considerably simplifies the continuum subtraction. By equating now $\Pi ^{%
\mathrm{OPE}}(M^{2})$ with the expression Eq.\ (\ref{eq:CorrF5}) and
performing the continuum subtraction, we find the sum rule equality which
depends on $\Pi ^{\mathrm{OPE}}(M^{2},s_{0})$: After the subtraction
procedure the correlation function $\Pi ^{\mathrm{OPE}}(M^{2},s_{0})$
acquires dependence on the continuum threshold parameter $s_{0}$. The
formulas necessary to carry out subtractions can be found in Appendix B of
Ref.\ \cite{Agaev:2016srl}.

By equating invariant amplitudes corresponding to aforementioned Lorentz
structures in both sides of the sum rule equality, one finds four equations
which should be solved to determine sum rules for the strong couplings. We
denote invariant amplitudes corresponding to the structures $\slashed q%
\slashed p\gamma _{5}$, $\slashed p\gamma _{5}$, $\slashed q\gamma _{5}$ and
$\gamma _{5}$ by $\Pi _{i}^{\mathrm{OPE}}(M^{2},s_{0})$, where $i=1,2,3$ and
$4$, respectively.

The solution of these equations for the coupling of interest $g_{1}$ is
\begin{eqnarray}
&&g_{1}=\frac{e^{\widetilde{m}^{2}/M_{1}^{2}}e^{m_{\Lambda }^{2}/M_{2}^{2}}}{%
\widetilde{\lambda }\lambda _{\Lambda }(m+\widetilde{m})(m_{\Lambda }+%
\widetilde{m}_{\Lambda })}\left\{ \Pi _{1}^{\mathrm{OPE}}\left[
m_{K}^{2}\right. \right.  \notag \\
&&\left. +m(\widetilde{m}_{\Lambda }-\widetilde{m})\right] +\Pi _{2}^{%
\mathrm{OPE}}(\widetilde{m}-m-\widetilde{m}_{\Lambda })  \notag \\
&&\left. +\Pi _{3}^{\mathrm{OPE}}(\widetilde{m}_{\Lambda }-\widetilde{m}%
)-\Pi _{4}^{\mathrm{OPE}}\right\} .  \label{eq:Coup1}
\end{eqnarray}%
Here, $m_{K}=(493.677\pm 0.016)~\mathrm{MeV}$ is the mass of the $K$ meson.
The sum rules for the strong coupling $g_{2}$ corresponding to the vertex $%
\Xi _{c}^{\prime }(1/2^{+})\Lambda _{c}^{+}K^{-}$ and responsible for the
decay $\Xi _{c}^{\prime }(1/2^{+})\rightarrow \Lambda _{c}^{+}K^{-}$ can be
determined from Eq.\ (\ref{eq:Coup1}) by replacements $\widetilde{m}%
\rightarrow -m^{\prime }$ and $\widetilde{\lambda }\rightarrow \lambda
^{\prime }$. \

In order to activate Eq.\ (\ref{eq:Coup1}), it is necessary to calculate the
correlation function $\Pi ^{\mathrm{OPE}}(p,q)$ and find the invariant
amplitudes $\Pi _{i}^{\mathrm{OPE}}(M^{2},s_{0})$. After contracting the
quarks fields and inserting into the obtained formula quark propagators, we
get the expression which depend on the non-local matrix elements of
operators $\overline{s}^{a}u^{b}$ placed between the states $\langle K(q)|$
and $|0\rangle $. We should express the correlation function $\Pi ^{\mathrm{%
OPE}}(p,q)$ using the distribution amplitudes (DAs) of $K$ meson with
different quark-gluon compositions and twists. To this end, we use the
expansion
\begin{equation}
\overline{s}_{\alpha }^{a}u_{\beta }^{b}=\frac{1}{12}\Gamma _{\beta \alpha
}^{i}\delta _{ab}(\overline{s}\Gamma ^{i}u),
\end{equation}%
where $\ \Gamma ^{i}=1,\ \gamma _{5},\ \gamma _{\mu },\ i\gamma _{5}\gamma
_{\mu },\ \sigma _{\mu \nu }/\sqrt{2}$ are the Dirac matrices. These terms
placed between the $K$ meson and vacuum states generate the two-particle DAs
of the leading and nonleading twists. They are defined by the expressions
\cite{Ball:2006wn}%
\begin{eqnarray}
&&\langle 0|\overline{q}(x)\gamma _{\mu }\gamma _{5}s(-x)|K(q)\rangle
=if_{K}q_{\mu }\int_{0}^{1}due^{i\xi qx}\left[ \phi _{2:K}(u)\right.  \notag
\\
&&\left. +\frac{1}{4}x^{2}\phi _{4:K}(u)\right] +\frac{i}{2}f_{K}\frac{%
x_{\mu }}{qx}\int_{0}^{1}due^{i\xi qx}\psi _{4:K}(u),  \label{eq:DA1}
\end{eqnarray}%
\begin{equation}
\langle 0|\overline{q}(x)i\gamma _{5}s(-x)|K(q)\rangle =\frac{f_{K}m_{K}^{2}%
}{m_{s}+m_{q}}\int_{0}^{1}due^{i\xi qx}\phi _{3:K}^{p}(u),  \label{eq:DA2}
\end{equation}%
and
\begin{eqnarray}
&&\langle 0|\overline{q}(x)\sigma _{\alpha \beta }\gamma
_{5}s(-x)|K(q)\rangle =-\frac{i}{3}\frac{f_{K}m_{K}^{2}}{m_{s}+m_{q}}  \notag
\\
&&\times (q_{\alpha }x_{\beta }-q_{\beta }x_{\alpha })\int_{0}^{1}due^{i\xi
qx}\phi _{3:K}^{\sigma }(u),  \label{eq:DA3}
\end{eqnarray}%
where $f_{K}=(155.72\pm 0.51)~\mathrm{MeV}$ is the decay constant of the $K$
meson. In expressions above $\xi =2u-1$, with $u$ being the longitudinal
momentum fraction carrying the quark in the $K$ meson. The subscripts in DAs
label the twist of these functions.
\begin{figure}[h]
\begin{center}
\includegraphics[totalheight=6cm,width=8cm]{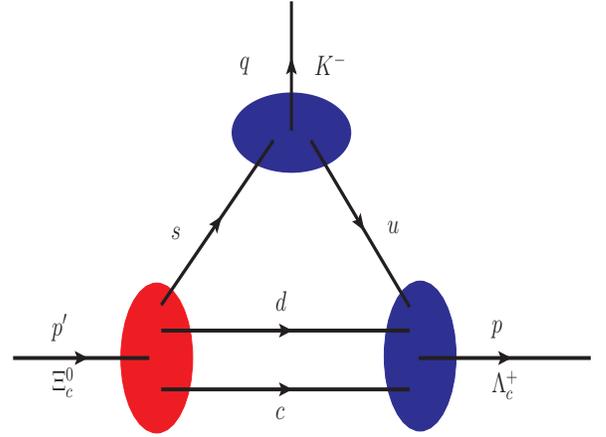}
\end{center}
\caption{The leading twist diagram contributing to $\Pi ^{\mathrm{OPE}}(p,q)$%
. }
\label{fig:Diag1}
\end{figure}

There are also three-particle twist-3 and -4 DAs of the kaon, which appear
due to insertions into operators $\overline{s}\Gamma ^{i}u$ of the gluon
field strength tensor $G_{\lambda \rho }$ coming from quark propagators. The
definitions of these DAs and their models are collected in Ref.\ \cite%
{Ball:2006wn}. The main contribution to $\Pi ^{\mathrm{OPE}}(p,q)$ arises
from the terms, where all the propagators are replaced by their perturbative
components. It is known as the leading twist contribution: the corresponding
Feynman diagram is plotted in Fig.\ \ref{fig:Diag1}. Contributions of terms
containing three-particle DA of the $K$ meson generate only nonleading twist
effects. In the present work, we take into account contributions due to two-
and three-particle DAs including twist-4 corrections. An analytic expression
for the double Borel transformed and subtracted correlation function $\Pi ^{%
\mathrm{OPE}}(M^{2},s_{0})$ is rather cumbersome, therefore we do not write
down it here. From derived expression of $\Pi ^{\mathrm{OPE}}(M^{2},s_{0})$
one can extract invariant amplitudes required for our calculations.

The functions $\Pi _{i}^{\mathrm{OPE}}(M^{2},s_{0})$ depend on DAs of $K$
meson. In numerical computations for these DAs, we have utilized models and
parameters presented in Ref. \cite{Ball:2006wn}. Apart from DAs, the sum
rules for the couplings $g_{1}$ and $g_{2}$ contain also masses of the
ground state $\Lambda _{c}^{+}$ and first orbitally excited $\Lambda
_{c}^{-} $ baryons for which we use their values from Ref.\ \cite%
{Tanabashi:2018oca}
\begin{eqnarray}
m_{\Lambda } &=&(2286.46\pm 0.14)\ \mathrm{MeV},\   \notag \\
\widetilde{m}_{\Lambda } &=&(2592.25\pm 0.28)\ \mathrm{MeV.}
\label{eq:Lambda}
\end{eqnarray}%
The pole residue of $\Lambda _{c}^{+}$ denoted in Eq.\ (\ref{eq:Coup1}) by $%
\lambda _{\Lambda }$ is borrowed from the work \cite{Aliev:2018ube}
\begin{equation}
\lambda _{\Lambda }=(3.8\pm 0.9)\times 10^{-2}\ \mathrm{GeV}^{3}.
\end{equation}%
The Borel and continuum threshold parameters for the decay of the baryons $%
\Xi _{c}^{\prime }(1/2^{+})$ and $\Xi _{c}^{\prime }(1/2^{+})$ are fixed
exactly as in computations of their masses. The parameters $\beta $ and $%
\widetilde{\beta }$ in the interpolating currents of $\Xi _{c}^{\prime }$
and $\Lambda _{c}$ are taken equal to each other and varied within the
limits presented in Eq.\ (\ref{eq:Beta}).

Numerical calculations lead to the following predictions
\begin{equation}
g_{1}=0.41\pm 0.04,\ |g_{2}|=7.19\pm 0.65.  \label{eq:Couplings}
\end{equation}%
The widths of the decays $\Xi _{c}^{\prime }(1/2^{-})\rightarrow \Lambda
_{c}^{+}K^{-}$ and $\Xi _{c}^{\prime }(1/2^{+})\rightarrow \Lambda
_{c}^{+}K^{-}$ can be obtained in terms of the strong couplings $g_{1}$ and $%
g_{2}$, respectively. They are determined by the formulas
\begin{eqnarray}
\Gamma \left( \Xi _{c}^{\prime }(1/2^{-})\rightarrow \Lambda
_{c}^{+}K^{-}\right) &=&\frac{g_{1}^{2}}{8\pi \widetilde{m}^{2}}\left[ (%
\widetilde{m}+m_{\Lambda })^{2}-m_{K}^{2}\right]  \notag \\
&&\times f(\widetilde{m},m_{\Lambda },m_{K}),  \label{eq:DW1}
\end{eqnarray}%
and%
\begin{eqnarray}
\Gamma \left( \Xi _{c}^{\prime }(1/2^{+})\rightarrow \Lambda
_{c}^{+}K^{-}\right) &=&\frac{g_{2}^{2}}{8\pi m^{\prime 2}}\left[ (m^{\prime
}-m_{\Lambda })^{2}-m_{K}^{2}\right]  \notag \\
&&\times f(m^{\prime },m_{\Lambda },m_{K}),  \label{eq:DW2}
\end{eqnarray}%
where the function $f(x,y,z)$ is given by the expression
\begin{eqnarray}
f(x,y,z) &=&\frac{1}{2x}\sqrt{%
x^{4}+y^{4}+z^{4}-2x^{2}y^{2}-2x^{2}z^{2}-2y^{2}z^{2}}.  \notag \\
&&
\end{eqnarray}%
The predictions for the width of the decays $\Xi _{c}^{\prime
}(1/2^{-})\rightarrow \Lambda _{c}^{+}K^{-}$ and $\Xi _{c}^{\prime
}(1/2^{+})\rightarrow \Lambda _{c}^{+}K^{-}$ are equal to
\begin{eqnarray}
\Gamma \left( \Xi _{c}^{\prime }(1/2^{-})\rightarrow \Lambda
_{c}^{+}K^{-}\right) &=&(7.3\pm 1.4)~\mathrm{MeV},  \notag \\
\Gamma \left( \Xi _{c}^{\prime }(1/2^{+})\rightarrow \Lambda
_{c}^{+}K^{-}\right) &=&(14.2\pm 2.7)~\mathrm{MeV}.  \label{eq:R1}
\end{eqnarray}%
Theoretical ambiguities in Eq.\ (\ref{eq:R1}) are generated by the strong
couplings $g_{1}^{2}$ and $g_{2}^{2}$, and by the masses $\widetilde{m}$ and
$m^{\prime }$of excited baryons $\Xi _{c}^{\prime }(1/2^{-})$ and $\Xi
_{c}^{\prime }(1/2^{+})$, which have been extracted in the present work. For
the masses $m_{\Lambda }$, $\widetilde{m}_{\Lambda }$ and $m_{K}$, we use
their experimental values, which are known with high precision: Relevant
experimental errors are very small and do not affect error estimates for
decay widths.

Predictions for partial widths of these two channels can be used for further
studies of the baryons $\Xi _{c}^{\prime }(1/2^{-})$ and $\Xi _{c}^{\prime
}(1/2^{+})$.


\section{Processes $\Xi _{c}^{\prime }(1/2^{-})\rightarrow \Xi _{c}^{\prime }%
\protect\pi $ and $\Xi _{c}^{\prime }(1/2^{+})\rightarrow \Xi _{c}^{\prime }%
\protect\pi $}

\label{sec:Decays1/2Pi}
Analysis of the decays $\Xi _{c}^{\prime }(1/2^{-})\rightarrow \Xi
_{c}^{\prime }\pi $ and $\Xi _{c}^{\prime }(1/2^{+})$ $\rightarrow \Xi
_{c}^{\prime }\pi $ does not differ from our studies carried out in the
previous section. The correlation function to investigate these processes is
given by the expression
\begin{equation}
\Pi _{\pi }(p,q)=i\int d^{4}xe^{ipx}\langle \pi (q)|\mathcal{T}\{\eta (x)%
\overline{\eta }(0)\}|0\rangle ,  \label{eq:CorrF3a}
\end{equation}%
with the same interpolating current\ $\eta (x)$.

Let us consider the decay $\Xi _{c}^{\prime }(1/2^{-})\rightarrow \Xi
_{c}^{\prime }\pi $ to outline methods of analysis. For this process the
physical side of the LCSR has the form
\begin{eqnarray}
\Pi _{\pi }^{\mathrm{Phys}}(p,q) &=&\frac{\langle 0|\eta |\Xi _{c}^{\prime
}(p,s)\rangle }{p^{2}-m^{2}}\langle \pi (q)\Xi _{c}^{\prime }(p,s)|\Xi
_{c}^{\prime }(1/2^{-},p^{\prime },s^{\prime })\rangle  \notag \\
&&\times \frac{\langle \Xi _{c}^{\prime }(1/2^{-},p^{\prime },s^{\prime })|%
\overline{\eta }|0\rangle }{p^{\prime 2}-\widetilde{m}^{2}}+\cdots .
\label{eq:CorrF3b}
\end{eqnarray}%
Matrix elements of the ground-state and excited baryons $\Xi _{c}^{\prime }$
and $\Xi _{c}^{\prime }(1/2^{-})$ are well known. Additionally, we model the
vertex matrix element by introducing the strong coupling $g_{3}$
\begin{equation}
\langle \pi (q)\Xi _{c}^{\prime }(p,s)|\Xi _{c}^{\prime }(1/2^{-},p^{\prime
},s^{\prime })\rangle =g_{3}\overline{u}(p,s)u(p^{\prime },s^{\prime }).
\end{equation}%
Then, the double Borel transformation of the correlation function is
determined by the expression
\begin{eqnarray}
&&\mathcal{B}\Pi _{\pi }^{\mathrm{Phys}}(p^{2},p^{\prime 2})=g_{3}\widetilde{%
\lambda }\lambda e^{-\widetilde{m}^{2}/M_{1}^{2}}e^{-m^{2}/M_{2}^{2}}  \notag
\\
&&\times (\slashed p+m)\gamma _{5}\left( \slashed p^{\prime }-\widetilde{m}%
\right).  \label{eq:CorrF3c}
\end{eqnarray}

The QCD side of the LCSR is determined by the correlator $\Pi _{\pi }(p,q)$
expressed in terms of the $c$ and $s$ quark propagators and distribution
amplitudes of the pion. The matrix elements of operators $\overline{d}%
(x)\Gamma ^{j}d(0)$ in $\Pi _{\pi }^{\mathrm{OPE}}(p,q)$ can be expanded
over $x^{2}$ and expressed by means of the pion's two-particle DAs of
different twist \cite{Braun:1989iv,Ball:1998je}. As examples, in the case of
$\Gamma =\ i\gamma _{\mu }\gamma _{5}$ and $\gamma _{5}$ we get
\begin{eqnarray}
&&\sqrt{2}\langle \pi ^{0}(q)|\overline{d}(x)i\gamma _{\mu }\gamma
_{5}d(0)|0\rangle  \notag \\
&=&f_{\pi }q_{\mu }\int_{0}^{1}due^{i\overline{u}qx}\left[ \phi _{\pi }(u)+%
\frac{m_{\pi }^{2}x^{2}}{16}\mathbb{A}_{4}(u)\right]  \notag \\
&&+\frac{f_{\pi }m_{\pi }^{2}}{2}\frac{x_{\mu }}{qx}\int_{0}^{1}due^{i%
\overline{u}qx}\mathbb{B}_{4}(u),  \label{eq:LTDA}
\end{eqnarray}%
and%
\begin{eqnarray}
&&2\sqrt{2}m_{d}\langle \pi ^{0}(q)|\overline{d}(x)i\gamma _{5}d(0)|0\rangle
=  \notag \\
&&\times f_{\pi }m_{\pi }^{2}\int_{0}^{1}due^{iuqx}\phi _{3;\pi }^{p}(u),
\label{eq:TW3}
\end{eqnarray}%
where $m_{\pi }=135~\mathrm{MeV}$ and $f_{\pi }=131~\mathrm{MeV}$ are
parameters of the pion $\pi ^{0}$. In Eq.\ (\ref{eq:LTDA}) $\phi _{\pi }(u)$
is the twist-2 distribution amplitude, and $\mathbb{A}_{4}(u)$ and $\mathbb{B%
}_{4}(u)$ are higher-twist functions that can be expressed in terms of the
pion two-particle twist-4 DAs. One of two-particle twist-3 distributions $%
\phi _{3;\pi }^{p}(u)$ determines the matrix element given by Eq.\ (\ref%
{eq:TW3}). Another twist-3 DA $\phi _{3;\pi }^{\sigma }(u)$ corresponds to
matrix element with $\sigma _{\mu \nu }$ insertion. The non-local operators
that appear due to a gluon field strength tensor $G_{\mu \nu }(ux)$
insertions to $\overline{d}(x)\Gamma ^{j}d(0)$ generate the pion's
three-particle distributions. Their definitions and further details were
presented in Ref.\ \cite{Ball:2006wn,Braun:1988qv,Braun:1989iv,Ball:1998je}.

The leading twist DA $\phi _{\pi }(u)$ is expressible in terms of the
Gegenbauer polynomials $C_{2n}^{3/2}(\varsigma )$
\begin{equation}
\phi _{\pi }(u,\mu ^{2})=6u\overline{u}\left[ 1+\sum_{n=1,2,3\ldots
}a_{2n}(\mu ^{2})C_{2n}^{3/2}(u-\overline{u})\right] ,  \label{eq:PionDALT}
\end{equation}%
where $\overline{u}=1-u$. The coefficients $a_{2n}(\mu ^{2})$ depend on a
scale $\mu $, as a result, $\phi _{\pi }(u,\mu ^{2})$ is a function of $\mu $%
. The Gegenbauer moments $a_{2n}(\mu ^{2})$ at some normalization point $\mu
=\mu _{0}$ should be fixed from phenomenological analysis or computed by
employing, for example, lattice simulations.

We derive the sum rule for the coupling $g_{3}$ by employing invariant
amplitudes corresponding to the structure $\slashed q\slashed p\gamma _{5}$
both in $\Pi _{\pi }^{\mathrm{Phys}}(p,q)$ and $\Pi _{\pi }^{\mathrm{OPE}%
}(p,q)$, and get
\begin{equation}
g_{3}=\frac{e^{\widetilde{m}^{2}/M_{1}^{2}}e^{m^{2}/M_{2}^{2}}}{\widetilde{%
\lambda }\lambda }\Pi _{\pi }^{\mathrm{OPE}},  \label{eq:XiPi}
\end{equation}%
where $\Pi _{\pi }^{\mathrm{OPE}}$ is the relevant amplitude. In
computations the Borel and continuum threshold parameters $M^{2}$ and $s_{0}$
are chosen as in Table\ \ref{tab:Results1A}. Parameters of the pion's DAs
are borrowed from Refs.\ \cite{Braun:1988qv,Braun:1989iv}. Numerical
calculations of $g_{3}$ and the partial width of the decay $\Xi _{c}^{\prime
}(1/2^{-})\rightarrow \Xi _{c}^{\prime }\pi $ lead to results%
\begin{equation}
g_{3}=0.14\pm 0.02,  \label{eq:Ncoupl1}
\end{equation}%
and
\begin{equation}
\Gamma \left( \Xi _{c}^{\prime }(1/2^{-})\rightarrow \Xi _{c}^{\prime }\pi
\right) =(1.1\pm 0.2)~\mathrm{MeV}.  \label{eq:Ndecay1}
\end{equation}%
The similar analysis can be done in the case of the second process $\Xi
_{c}^{\prime }(1/2^{+})\rightarrow \Xi _{c}^{\prime }\pi $. The strong
coupling $g_{4}$ can be obtained from Eq.\ (\ref{eq:XiPi}) by replacements $%
\widetilde{m}\rightarrow m^{\prime }$ and $\widetilde{\lambda }\rightarrow
\lambda ^{\prime }$. \ Computations give following predictions
\begin{eqnarray}
&&g_{4}=1.04\pm 0.16,  \notag \\
&&\Gamma \left( \Xi _{c}^{\prime }(1/2^{+})\rightarrow \Xi _{c}^{\prime }\pi
\right) =(0.39\pm 0.09)~\mathrm{MeV}.  \label{eq:Ndecay2}
\end{eqnarray}%
Information obtained here will be used to evaluate widths of the baryons $%
\Xi _{c}^{\prime }(1/2^{-})$ and $\Xi _{c}^{\prime }(1/2^{+})$.


\section{ Decay channels $\Xi _{c}^{\ast }(3/2^{-})\rightarrow \Lambda
_{c}^{+}K^{-},\Xi _{c}^{\prime }\protect\pi $ and $\Xi _{c}^{^{\ast
}}(3/2^{+})\rightarrow \Lambda _{c}^{+}K^{-},\Xi _{c}^{\prime }\protect\pi $}

\label{sec:Decays3/2}

The decays of the spin-$3/2$ baryons $\Xi _{c}^{\ast }(3/2^{-})$ and $\Xi
_{c}^{^{\ast }}(3/2^{+})$ to the final state $\Lambda _{c}^{+}\,K^{-}$ can
be explored by a manner described above for the spin-$1/2$ particles. \ To
this end, we begin from calculation of the correlation function
\begin{equation}
\Pi _{\mu }(p,q)=i\int d^{4}xe^{ipx}\langle K(q)|\mathcal{T}\{\eta _{\Lambda
}(x)\overline{\eta }_{\mu }(0)\}|0\rangle ,  \label{eq:CorrF6}
\end{equation}%
where $\eta _{\mu }$ is the interpolating current for spin-$3/2$ baryons $%
\Xi _{c}^{\ast }$ given by Eq.\ (\ref{eq:BayC3/2}).

To calculate the phenomenological side of the sum rules $\Pi _{\mu }^{%
\mathrm{Phys}}(p,q)$, we write down it in the form similar to one presented
in Eq.\ (\ref{eq:CorrF4}) with simple modifications. We also define the
strong couplings $G_{0(1)}$ and $\widetilde{G}_{0(1)}$ using the matrix
elements
\begin{eqnarray}
&&\langle K(q)\Lambda _{c}^{+}(p,s)|\Xi _{c}^{\ast }(p^{\prime },s^{\prime
})\rangle =G_{0}\overline{u}(p,s)u_{\alpha }(p^{\prime },s^{\prime
})q^{\alpha },  \notag \\
&&\langle K(q)\Lambda _{c}^{-}(p,s)|\Xi _{c}^{\ast }(p^{\prime },s^{\prime
})\rangle =\widetilde{G}_{0}\overline{u}(p,s)\gamma _{5}u_{\alpha
}(p^{\prime },s^{\prime })q^{\alpha },  \notag \\
&&\langle K(q)\Lambda _{c}^{+}(p,s)|\Xi _{c}^{\ast }(3/2^{-},p^{\prime
},s^{\prime })\rangle =G_{1}\overline{u}(p,s)\gamma _{5}u_{\alpha
}(p^{\prime },s^{\prime })q^{\alpha },  \notag \\
&&{}\langle K(q)\Lambda _{c}^{-}(p,s)|\Xi _{c}^{\ast }(3/2^{-},p^{\prime
},s^{\prime })\rangle =\widetilde{G}_{1}\overline{u}(p,s)u_{\alpha
}(p^{\prime },s^{\prime })q^{\alpha }.  \notag \\
&&  \label{Eq.ME5}
\end{eqnarray}%
After some manipulations, for the Borel transformation of $\Pi _{\mu }^{%
\mathrm{Phys}}(p,q)$, we obtain the following expression
\begin{eqnarray}
&&\mathcal{B}\Pi _{\mu }^{\mathrm{Phys}}(p^{2},p^{\prime 2})=G_{0}\lambda
^{\ast }\lambda _{\Lambda }e^{-m^{\ast 2}/M_{1}^{2}}e^{-m_{\Lambda
}^{2}/M_{2}^{2}}(\slashed p+m_{\Lambda })  \notag \\
&&\times \left( \slashed p^{\prime }+m^{\ast }\right) F_{\alpha \mu
}(m^{\ast },p^{\prime })q^{\alpha }-\widetilde{G}_{0}\lambda ^{\ast }%
\widetilde{\lambda }_{\Lambda }e^{-m^{\ast 2}/M_{1}^{2}}  \notag \\
&&\times e^{-\widetilde{m}_{\Lambda }^{2}/M_{2}^{2}}(\slashed p-\widetilde{m}%
_{\Lambda })\left( \slashed p^{\prime }+m^{\ast }\right) F_{\alpha \mu
}(m^{\ast },p^{\prime })q^{\alpha }  \notag \\
&&+G_{1}\widetilde{\lambda }^{\ast }\lambda _{\Lambda }e^{-\widetilde{m}%
^{\ast 2}/M_{1}^{2}}e^{-m_{\Lambda }^{2}/M_{2}^{2}}(\slashed p+m_{\Lambda
})\left( \slashed p^{\prime }-\widetilde{m}^{\ast }\right) \gamma _{5}
\notag \\
&&\times F_{\alpha \mu }(\widetilde{m}^{\ast },p^{\prime })\gamma
_{5}q^{\alpha }-\widetilde{G}_{1}\widetilde{\lambda }^{\ast }\widetilde{%
\lambda }_{\Lambda }e^{-\widetilde{m}^{\ast 2}/M_{1}^{2}}e^{-\widetilde{m}%
_{\Lambda }^{2}/M_{2}^{2}}  \notag \\
&&\times (\slashed p-\widetilde{m}_{\Lambda })\left( \slashed p^{\prime }-%
\widetilde{m}^{\ast }\right) \gamma _{5}F_{\alpha \mu }(\widetilde{m}^{\ast
},p^{\prime })\gamma _{5}q^{\alpha }.  \label{eq:CorrF7}
\end{eqnarray}%
To derive the sum rules, we use available structures in Eq.\ (\ref{eq:CorrF7}%
). The same terms are fixed in $\mathcal{B}\Pi _{\mu }^{\mathrm{QCD}%
}(p^{2},p^{\prime 2})$ and matched with ones from $\mathcal{B}\Pi _{\mu }^{%
\mathrm{Phys}}(p^{2},p^{\prime 2})$. The final expressions \ of the strong
couplings are rather lengthy, therefore we do not write down them here.

The strong coupling required to compute the width of the decay $\Xi
_{c}^{\ast }(3/2^{-})\rightarrow \Lambda _{c}^{+}K^{-}$ is $G_{1}$. The
coupling $G_{2}$ necessary to find the width of the process $\Xi _{c}^{\ast
}(3/2^{+})\rightarrow \Lambda _{c}^{+}K^{-}$ can be obtained from the
relevant sum rule after simple replacements. In numerical computations the
parameters $M^{2}$, $s_{0}$ are chosen as in the corresponding mass
calculations. For $G_{1}$ and $G_{2}$ our analysis leads to the following
predictions (in units of $\mathrm{GeV}^{-1}$)
\begin{equation}
G_{1}=22.0\pm 2.6,\ \ |G_{2}|=3.6\pm 0.4.  \label{eq:Couplings2}
\end{equation}

The information gained from these studies is enough to determine the widths
of the corresponding decay channels. In fact, the width of the decay $\Xi
_{c}^{\ast }(3/2^{-})\rightarrow \Lambda _{c}^{+}K^{-}$ can be found using
the expression%
\begin{eqnarray}
\Gamma (\Xi _{c}^{\ast }(3/2^{-}) &\rightarrow &\Lambda _{c}^{+}K^{-})=\frac{%
G_{1}^{2}}{24\pi \widetilde{m}^{\ast 2}}\left[ (\widetilde{m}^{\ast
}-m_{\Lambda })^{2}-m_{K}^{2}\right]  \notag \\
&&\times f^{3}(\widetilde{m}^{\ast },m_{\Lambda },m_{K}),
\end{eqnarray}%
whereas for $\Gamma (\Xi _{c}^{\ast }(3/2^{+})\rightarrow \Lambda
_{c}^{+}K^{-})$, we employ
\begin{eqnarray}
\Gamma (\Xi _{c}^{\ast }(3/2^{+}) &\rightarrow &\Lambda _{c}^{+}K^{-})=\frac{%
G_{2}^{2}}{24\pi m^{\ast \prime ^{2}}}\left[ (m^{\ast \prime }+m_{\Lambda
})^{2}-m_{K}^{2}\right]  \notag \\
&&\times f^{3}(m^{\ast \prime },m_{\Lambda },m_{K}).
\end{eqnarray}%
Numerical computations lead to predictions
\begin{eqnarray}
&&\Gamma \left( \Xi _{c}^{\ast }(3/2^{-})\rightarrow \Lambda
_{c}^{+}K^{-}\right) =(10.5\pm 2.1)~\mathrm{MeV},  \notag \\
&&\Gamma \left( \Xi _{c}^{\ast }(3/2^{+})\rightarrow \Lambda
_{c}^{+}K^{-}\right) =(35.4\pm 7.2)~\mathrm{MeV}.
\end{eqnarray}

Exploration of the second pair of processes $\Xi _{c}^{\ast
}(3/2^{-})\rightarrow \Xi _{c}^{\prime }\pi $ and $\Xi _{c}^{\ast
}(3/2^{+})\rightarrow \Xi _{c}^{\prime }\pi $ is performed using the
correlation function

\begin{equation}
\Pi _{\mu }^{\pi }(p,q)=i\int d^{4}xe^{ipx}\langle \pi (q)|\mathcal{T}\{\eta
(x)\overline{\eta }_{\mu }(0)\}|0\rangle .  \label{eq:CorrF6a}
\end{equation}%
The sum rules for $G_{3}$ and $G_{4}$ are obtained by means of the
structures $\slashed q\slashed pq_{\mu }$ in the physical and QCD
representation of the correlator $\Pi _{\mu }^{\pi }(p,q)$. Our analysis for
strong couplings and partial widths yields:%
\begin{eqnarray}
&&G_{3}=4.3\pm 0.6,  \notag \\
&&\Gamma \left( \Xi _{c}^{\ast }(3/2^{-})\rightarrow \Xi _{c}^{\prime }\pi
\right) =(4.0\pm 0.8)\times 10^{-2}~\mathrm{MeV},  \notag \\
&&  \label{eq.NDecay3}
\end{eqnarray}%
and
\begin{eqnarray}
&&G_{4}=0.45\pm 0.06,  \notag \\
&&\Gamma \left( \Xi _{c}^{\ast }(3/2^{+})\rightarrow \Xi _{c}^{\prime }\pi
\right) =(1.7\pm 0.4)\times 10^{-1}~\mathrm{MeV}.  \notag \\
&&  \label{eq:NDecay4}
\end{eqnarray}

Partial widths of processes considered here allow us to evaluate the full
widths of the baryons $\Xi _{c}^{\ast }(3/2^{-})$ and $\Xi _{c}^{\ast
}(3/2^{+})$ saturated by two dominant decay modes. It is clear that effect
of the process $\Xi _{c}^{\ast }(3/2^{-})\rightarrow \Xi _{c}^{\prime }\pi $
on the full width of $\Xi _{c}^{\ast }(3/2^{-})$ is negligible, and can be
safely ignored. For width of the baryon $\Xi _{c}^{\ast }(3/2^{+})$, we get
\begin{equation}
\Gamma ^{\ast \prime }=(35.6\pm 2.1)~\mathrm{MeV}.  \label{eq:Results1}
\end{equation}%
This estimate for $\Gamma ^{\ast \prime }$, and prediction $\widetilde{%
\Gamma }^{\ast }=(10.5\pm 2.1)~\mathrm{MeV}$ should be compared with the
LHCb data.


\section{Analysis and Concluding notes}

\label{sec:Concl}

We have calculated the masses and pole residues of the ground state spin-$%
1/2 $ and -$3/2$ baryons $\Xi _{c}^{\prime }$ and $\Xi _{c}^{\ast }$
\begin{equation}
m_{\mathrm{th}}=(2576\pm 150)~\mathrm{MeV},
\end{equation}%
and
\begin{equation}
m_{\mathrm{th}}^{\ast }=(2655\pm 102)~\mathrm{MeV}.
\end{equation}%
Comparing obtained theoretical predictions for masses of these particles
with experimental data (\ref{eq:Data2}) and (\ref{eq:Data3}), we see nice
agreements between them: Theoretical errors of $m_{\mathrm{th}}$ and $m_{%
\mathrm{th}}^{\ast }$ are typical for this method and do not exceed allowed
limits.

In the present work we have also computed the masses and widths of the spin-$%
1/2$ flavor-sextet baryons $\Xi _{c}^{\prime }$, and spin-$3/2$ baryons $\Xi
_{c}^{\ast }$ in order to compare obtained information with results of the
LHCb collaboration. The masses of these particles have been extracted from
two-point sum rules, whereas to calculate their widths, we have used the QCD
light-cone sum rule approach.

The sum rule method is a powerful nonperturbative tool to explore features
of conventional and exotic hadrons. It relies on first principles on the QCD
by employing quark-gluon structure of particles under analysis, and
universal vacuum expectations values of various local quark, gluon, and
mixed operators. Predictions obtained in this context depend on a few
auxiliary parameters of computations, which limit theoretical accuracy of
investigations. Main part of uncertainties is generated by a choice of the
Borel parameter $M^{2}$: its variation within allowed working region leads
to ambiguities in values of extracted parameters. In this sense the mass of
a hadron is most protected physical quantity the reason being in a
functional form of a relevant sum rule. In fact, sum rules for the masses of
hadrons are given as a ratio of correlation functions (see, for instance
Eq.\ (\ref{eq:SR1})), which reduces uncertainties and stabilize a final
result.

In the present article ambiguities in the masses of the excited spin-$1/2$
and -$3/2$ baryons $\Xi _{c}^{\prime }$ and $\Xi _{c}^{\ast }$ amount to $%
\pm (2.2-3.8)\%$ of central values, which is nice accuracy for sum rule
computations. In other words, the masses of the baryons may be chosen from
values spanning approximately $(120-220)~\mathrm{MeV}$ region. Because,
resonances discovered by LHCb have very close masses and cover narrow range
of $\sim 40~\mathrm{MeV}$, the sum rule method could not resolve such fine
structure: its predictions are compatible with all of these resonances.
Performed analysis also does not "see" resonances $\Xi _{c}(2790)$ and $\Xi
_{c}^{\ast }(2815)$, because central values for masses of the excited states
extracted in the present work are higher than masses of these particles.
Therefore, classification of the spin-$1/2$ and -$3/2$ excited baryons $\Xi
_{c}^{\prime }$ and $\Xi _{c}^{\ast }$, and their possible interpretation as
resonances $\Xi _{c}(2923)^{0}$, $\Xi _{c}(2939)^{0}$, and $\Xi
_{c}(2965)^{0}$ should be done using widths of these particles, which differ
from each other and have been evaluated with accuracy enough for such
differentiation.

Let us note that parameters of the flavor-antitriplet spin-$1/2$ states $csd$
were calculated in Ref.\ \cite{Aliev:2018ube}. In that paper the authors
considered the baryon $\Xi _{c}(1/2^{-})$ with parameters $\widetilde{m}%
=(2922\pm 83)~\mathrm{MeV}$, and $\widetilde{\Gamma }=(19.4\pm 3.3)~\mathrm{%
MeV}$ as the resonance $\Xi _{c}(2930)^{0}$. The radial excitation of the
spin-$1/2$ antitriplet baryon $\Xi _{c}(1/2^{+})$ has the same mass but
lower width
\begin{eqnarray}
m^{\prime } &=&(2922\pm 83)~\mathrm{MeV},  \notag \\
\Gamma ^{\prime } &=&(13.6\pm 2.3)~\mathrm{MeV}.  \label{eq:AntiTr2S}
\end{eqnarray}%
This particle should be taken into account in our present analysis.

Before confronting with experimental data the full widths of the
flavor-sextet baryons should be found using results of the sections \ref%
{sec:Decays1/2} and \ref{sec:Decays1/2Pi}. Simple analysis leads to the
following estimates
\begin{equation}
\widetilde{\Gamma }=(8.4\pm 1.4)\mathrm{~MeV},\ \Gamma ^{\prime }=(14.6\pm
2.7)\mathrm{~MeV.}  \label{eqFW1/2}
\end{equation}%
Then, it is not difficult to see that sextet baryon $\Xi _{c}^{\prime
}(1/2^{-})$, which has the mass and width
\begin{eqnarray}
\widetilde{m} &=&(2925\pm 115)~\mathrm{MeV},  \notag \\
\widetilde{\Gamma } &=&(8.4\pm 1.4)~\mathrm{MeV},\
\end{eqnarray}%
can be interpreted as the resonance $\Xi _{c}(2923)^{0}$ with parameters (%
\ref{eq:Data5}).

The second resonance $\Xi _{c}(2939)^{0}$ may be considered as the excited
spin-$3/2$ baryon $\Xi _{c}^{\ast }(3/2^{-})$
\begin{eqnarray}
\widetilde{m}^{\ast } &=&(2960\pm 67)~\mathrm{MeV},  \notag \\
\widetilde{\Gamma }^{\ast } &=&(10.5\pm 2.1)~\mathrm{MeV}.
\end{eqnarray}

The interpretation of $\Xi _{c}(2965)^{0}$ with parameters (\ref{eq:Data7})
is twofold: it may be considered as the spin-$1/2$ antitriplet baryon $\Xi
_{c}(1/2^{+})$ with $\Gamma ^{\prime }=(13.6\pm 2.3)~\mathrm{MeV}$. But one
can identify it also with radially excited sextet particle $\Xi _{c}^{\prime
}(1/2^{+})$ with the width $(14.2\pm 2.7)~\mathrm{MeV}$. Let us note that
masses of these particles within theoretical errors are compatible with the
LHCb data.

Because the radially excited spin-$3/2$ particle $\Xi _{c}^{\ast }(3/2^{+})$
has the width $(35.6\pm 2.1)~\mathrm{MeV}$, it cannot be considered as a
candidate to new resonances. It is worth noting that parameters of $\Xi
_{c}^{\ast }(3/2^{+})$ are close to the mass and width of the baryon $\Xi
_{c}(2970)^{0}$ \cite{Yelton:2016fqw}
\begin{eqnarray}
m &=&(2970.8\pm 0.7\pm 0.2)~\mathrm{MeV}  \notag \\
\Gamma &=&30.3\pm 2.3_{-4.0}^{+1.0}~\mathrm{MeV},
\end{eqnarray}%
Recently the Belle collaboration determined the spin-parity of $\Xi
_{c}(2970)$ as $J^{\mathrm{P}}=1/2^{+\text{ }}$\cite{Moon:2020gsg}.
Conclusion about radially excited spin-$1/2$ nature of the baryon $\Xi
_{c}(2970)$ was made also in Ref.\ \cite{Arifi:2020yfp}. In the light of
these circumstances $\Xi _{c}^{\ast }(3/2^{+})$ cannot be interpreted as $%
\Xi _{c}(2970)$. Identification of two resonances $\Xi _{c}(2965)^{0}$ and $%
\Xi _{c}(2970)^{0}$ seems also to be problematic, because they have
significantly different decay widths.

We have noted above that new LHCb resonances were explored in Refs.\ \cite%
{Yang:2020zjl,Wang:2020gkn,Lu:2020ivo,Zhu:2020jke} using various suggestions
on their structure and employing different computational schemes. Thus, in
Ref.\ \cite{Yang:2020zjl} these states were investigated in the context of
the HQET. In this theory a $P$-wave baryon consists of one charm quark and
light diquark, and contains one orbital excitation between light quarks ($%
\rho $-mode), or between a charm and light diquark ($\lambda $-mode). In
this paper, the resonances $\Xi _{c}(2923)^{0}$, $\Xi _{c}(2939)^{0}$, and $%
\Xi _{c}(2965)^{0}/\Xi _{c}(2970)^{0}$ were interpreted as $P$-wave baryons $%
J^{\mathrm{P}}=1/2^{-}$, $J^{\mathrm{P}}=3/2^{-}$ and $J^{\mathrm{P}%
}=3/2^{-} $containing one $\lambda $-mode: A difference in organization of
two last resonances with $J^{\mathrm{P}}=3/2^{-}$ was explained in Ref.\
\cite{Yang:2020zjl}. Our interpretation of $\Xi _{c}(2923)^{0}$, and $\Xi
_{c}(2939)^{0}$ is consistent with this picture. But in our analysis the
last resonance from this list $\Xi _{c}(2965)^{0}$ is either $2S$
antitriplet or sextet spin-$1/2$ particle.

The new resonances were considered also in Ref.\ \cite{Wang:2020gkn}, in
which first two states were interpreted as $\lambda $-mode baryons with $J^{%
\mathrm{P}}=3/2^{-}$, and the last state as $J^{\mathrm{P}}=5/2^{-}$.
Interpretation of only the resonance $\Xi _{c}(2939)^{0}$ in this scheme
coincides with our prediction. In Ref.\ \cite{Lu:2020ivo} the last particle $%
\Xi _{c}(2965)^{0}$ was regarded as spin-$1/2$ flavor-sextet $2S$ baryon,
which is in accord with our assignment.

New measurements by LHCb provided information on parameters of three
resonances which can be considered as charm-strange baryons. In the present
work we have calculated the masses and widths of four excited $csd$ baryons
with different spins. Theoretical investigations of first orbitally and
radially excited spin-$1/2$ sextet and spin-$3/2$ baryons, as well as
existing results for spin-$1/2$ flavor-antitriplet states fix parameters of
six particles. We have used\ new resonances $\Xi _{c}(2923)^{0}$, $\Xi
_{c}(2939)^{0}$, and $\Xi _{c}(2965)^{0}$, and known ones $\Xi
_{c}(2930)^{0} $ and $\Xi _{c}(2970)^{0}$ to confront their parameters with
our predictions: Obtained results have been discussed above. From brief
analysis of theoretical articles it is evident that interpretations of
excited $csd$ baryons are controversial. Evidently, for comprehensive
analysis of this sector of hadron spectroscopy more detailed experimental
information and further investigations are required.


\section*{ACKNOWLEDGEMENTS}

S.~S.~A. is grateful to Prof.~V.~M.~Braun for enlightening comments.

\appendix*

\section{ The invariant amplitudes $\Pi _{1}^{\mathrm{OPE}}(M^{2},s_{0})$
and $\Pi _{2}^{\mathrm{OPE}}(M^{2},s_{0})$}

\renewcommand{\theequation}{\Alph{section}.\arabic{equation}} \label{sec:App}
\begin{widetext}

The invariant amplitudes $\Pi _{1(2)}^{\mathrm{OPE}}(p^{2})$ used to
calculate the mass and pole residue of the flavor-sextet spin-$1/2$ baryons
after the Borel transformation and subtraction prescriptions take the
following form%
\begin{equation}
\Pi _{1(2)}^{\mathrm{OPE}}(M^{2},s_{0})=\int_{\mathcal{M}^{2}}^{s_{0}}ds\rho
_{1(2)}^{\mathrm{OPE}}(s)e^{-s/M^{2}}+\Pi _{1(2)}(M^{2}),  \label{eq:A0}
\end{equation}%
where $\mathcal{M}=m_{c}+m_{s}$. The spectral densities $\rho _{1(2)}^{%
\mathrm{OPE}}(s)$ in Eq.\ (\ref{eq:A0}) are found from the imaginary part of
the correlation function and encompass essential piece of $\Pi ^{\mathrm{OPE}%
}(p)$. The Borel transformations of remaining terms in $\Pi ^{\mathrm{OPE}%
}(p)$ are included into $\Pi _{1(2)}(M^{2})$, and have been calculated
directly from the expression of $\Pi ^{\mathrm{OPE}}(p)$.

The functions $\rho _{1(2)}^{\mathrm{OPE}}(s)$ and $\Pi _{1(2)}(M^{2})$
contain components of different dimensions and have the structure
\begin{equation}
\rho _{1(2)}^{\mathrm{OPE}}(s)=\rho _{1(2)}^{\mathrm{pert.}}(s)+\sum \rho
_{1(2)}^{\mathrm{DimN}}(s),\ \ \Pi _{1(2)}(M^{2})=\sum \Pi _{1(2)}^{\mathrm{%
DimN}}(M^{2}).  \label{eq:A1}
\end{equation}%
The $\Pi _{1(2)}^{\mathrm{OPE}}(M^{2},s_{0})$ have been computed by setting $%
m_{q}=0$ and $m_{s}\neq 0$, and used to perform numerical analyses. The
amplitudes $\Pi _{1(2)}^{\mathrm{OPE}}(M^{2},s_{0})$, in general, contain a
few hundred terms and exist as \textit{Mathematica} files. Their explicit
expressions are cumbersome, therefore we provide below simplified formulas
in which $m_{s}=0$.

The perturbative contribution and nonperturbative terms with dimensions $3$,
$4$, $5$ and $7$ in the case of the spectral density $\rho _{1}^{\mathrm{OPE}%
}(s)$ are given by the expressions:%
\begin{equation*}
\rho _{1}^{\mathrm{pert.}}(s)=\frac{(5+2\beta +5\beta ^{2})}{2048\pi
^{4}s^{2}}\left[
m_{c}^{6}(8s-m_{c}^{2})+s^{3}(s-8m_{c}^{2})+12m_{c}^{4}s^{2}\ln \left( \frac{%
s}{m_{c}^{2}}\right) \right] ,
\end{equation*}%
\begin{equation*}
\rho _{1}^{\mathrm{Dim3}}(s)=\frac{\langle \overline{s}s\rangle +\langle
\overline{d}d\rangle }{192\pi ^{2}s^{2}}m_{c}\left( m_{c}^{2}-s\right)
^{2}\left( 1+4\beta -5\beta ^{2}\right) ,
\end{equation*}%
\begin{equation*}
\rho _{1}^{\mathrm{Dim4}}(s)=\frac{\langle g_{s}^{2}G^{2}\rangle }{3072\pi
^{4}s^{2}}\left( s-m_{c}^{2}\right) \left[ 8m_{c}^{2}\left( 1+\beta +\beta
^{2}\right) +s\left( 5+2\beta +5\beta ^{2}\right) \right] ,
\end{equation*}%
\begin{equation*}
\rho _{1}^{\mathrm{Dim5}}(s)=\frac{\langle \overline{d}g_{s}\sigma Gd\rangle
+\langle \overline{s}g_{s}\sigma Gs\rangle }{768\pi ^{2}s^{2}}m_{c}(\beta -1)%
\left[ m_{c}^{2}(7+11\beta )-6s(1+\beta )\right] ,
\end{equation*}%
\begin{equation*}
\rho _{1}^{\mathrm{Dim7}}(s)=\frac{\langle g_{s}^{2}G^{2}\rangle \left[
\langle \overline{s}s\rangle +\langle \overline{d}d\rangle \right] }{384\pi
^{2}s^{2}}m_{c}(1-\beta ^{2}).
\end{equation*}%
The function $\Pi _{1}(M^{2})$ is composed of the following components:
\begin{equation*}
\Pi _{1}^{\mathrm{Dim6}}(M^{2},s_{0})=\frac{\langle \overline{s}s\rangle
\langle \overline{d}d\rangle }{72}\left( 11\beta ^{2}+2\beta -13\right)
e^{-m_{c}^{2}/M^{2}},
\end{equation*}%
\begin{equation*}
\Pi _{1}^{\mathrm{Dim7}}(M^{2},s_{0})=\frac{\langle g_{s}^{2}G^{2}\rangle %
\left[ \langle \overline{s}s\rangle +\langle \overline{d}d\rangle \right] }{%
864\pi ^{2}m_{c}}\left( \beta ^{2}+\beta -2\right) e^{-m_{c}^{2}/M^{2}},
\end{equation*}%
\begin{eqnarray*}
&&\Pi _{1}^{\mathrm{Dim8}}(M^{2},s_{0})=-\frac{\langle g_{s}^{2}G^{2}\rangle
^{2}}{27\cdot 2^{13}\pi ^{4}M^{2}}\left( 13\beta ^{2}+10\beta +13\right)
e^{-m_{c}^{2}/M^{2}}+\frac{\langle \overline{s}g_{s}\sigma Gs\rangle \langle
\overline{d}d\rangle }{288M^{4}}(1-\beta ) \\
&&\times \left[ m_{c}^{2}(26+22\beta )+M^{2}\left( 25+23\beta \right) \right]
e^{-m_{c}^{2}/M^{2}},
\end{eqnarray*}%
\begin{eqnarray*}
&&\Pi _{1}^{\mathrm{Dim9}}(M^{2},s_{0})=\frac{\langle \overline{s}%
g_{s}\sigma Gs\rangle \langle g_{s}^{2}G^{2}\rangle }{27\cdot 2^{11}\pi
^{2}m_{c}M^{4}}(1-\beta )\left[ m_{c}^{2}(31+11\beta )-2M^{2}(1+\beta )%
\right] e^{-m_{c}^{2}/M^{2}}, \\
&&\Pi _{1}^{\mathrm{Dim10}}(M^{2},s_{0})=0.
\end{eqnarray*}%
For the spectral density $\rho _{2}^{\mathrm{OPE}}(s)$, we get%
\begin{equation*}
\rho _{2}^{\mathrm{pert.}}(s)=\frac{m_{c}(13-2\beta -11\beta ^{2})}{1536\pi
^{4}s}\left[
m_{c}^{6}+9sm_{c}^{4}-9m_{c}^{2}s^{2}-s^{3}+6m_{c}^{2}s(s+m_{c}^{2})\ln
\left( \frac{s}{m_{c}^{2}}\right) \right] ,
\end{equation*}%
\begin{equation*}
\rho _{2}^{\mathrm{Dim3}}(s)=\frac{\langle \overline{s}s\rangle +\langle
\overline{d}d\rangle }{192\pi ^{2}s}\left( m_{c}^{2}-s\right) ^{2}\left(
1+4\beta -5\beta ^{2}\right) ,\
\end{equation*}%
\begin{eqnarray*}
&&\rho _{2}^{\mathrm{Dim4}}(s)=\frac{\langle g_{s}^{2}G^{2}\rangle }{9216\pi
^{4}m_{c}s}\left( 1-\beta \right) \left[ (m_{c}^{2}-s)\left( s(13+11\beta
)+m_{c}^{2}(53+67\beta )\right) \right.  \\
&&\left. +3m_{c}^{2}s(11+13\beta )\ln \left( \frac{s}{m_{c}^{2}}\right) %
\right] ,
\end{eqnarray*}%
\begin{equation*}
\rho _{2}^{\mathrm{Dim5}}(s)=\frac{\langle \overline{d}g_{s}\sigma Gd\rangle
+\langle \overline{s}g_{s}\sigma Gs\rangle }{768\pi ^{2}s}(1-\beta )\left[
m_{c}^{2}(5+\beta )-6s(1+\beta )\right] .
\end{equation*}%
The function $\Pi _{2}(M^{2})$ is determined by the components
\begin{equation*}
\Pi _{2}^{\mathrm{Dim6}}(M^{2},s_{0})=\frac{\langle \overline{s}s\rangle
\langle \overline{d}d\rangle }{24}m_{c}\left( 5\beta ^{2}+2\beta +5\right)
e^{-m_{c}^{2}/M^{2}},
\end{equation*}%
\begin{equation*}
\Pi _{2}^{\mathrm{Dim7}}(M^{2},s_{0})=\frac{\langle g_{s}^{2}G^{2}\rangle %
\left[ \langle \overline{s}s\rangle +\langle \overline{d}d\rangle \right] }{%
3456\pi ^{2}}\left( \beta ^{2}-8\beta +7\right) e^{-m_{c}^{2}/M^{2}},
\end{equation*}%
\begin{eqnarray*}
&&\Pi _{2}^{\mathrm{Dim8}}(M^{2},s_{0})=\frac{\langle g_{s}^{2}G^{2}\rangle
^{2}}{27\cdot 2^{13}\pi ^{4}m_{c}M^{2}}(m_{c}^{2}-2M^{2})\left( 11+2\beta
-13\beta ^{2}\right) e^{-m_{c}^{2}/M^{2}}+\frac{\langle \overline{s}%
g_{s}\sigma Gs\rangle \langle \overline{d}d\rangle }{144M^{4}}m_{c} \\
&&\times \left[ M^{2}\left( \beta -1\right) ^{2}-3m_{c}^{2}(5+2\beta +5\beta
^{2})\right] e^{-m_{c}^{2}/M^{2}},
\end{eqnarray*}%
\begin{eqnarray*}
&&\Pi _{2}^{\mathrm{Dim9}}(M^{2},s_{0})=\frac{\langle \overline{s}%
g_{s}\sigma Gs\rangle \langle g_{s}^{2}G^{2}\rangle }{27\cdot 2^{11}\pi
^{2}M^{4}}m_{c}^{2}(\beta ^{2}+28\beta -29)e^{-m_{c}^{2}/M^{2}}, \\
&&\Pi _{2}^{\mathrm{Dim10}}(M^{2},s_{0})=0.
\end{eqnarray*}

\end{widetext}

\end{document}